\documentclass[10pt,twocolumn,english]{aastex6}
\usepackage[T1]{fontenc}
\usepackage[latin9]{inputenc}
\usepackage{float}
\usepackage{units}
\usepackage{textcomp}
\usepackage{amsbsy}
\usepackage{amstext}
\usepackage{amssymb}
\usepackage{graphicx}
\usepackage[nointegrals]{wasysym}
\usepackage[bottom]{footmisc}

\makeatletter
\usepackage{natbib}
\usepackage{amsmath}
\setcitestyle{round}
\DeclareMathOperator{\sgn}{sgn}
\DeclareMathOperator{\Ai}{Ai}
\DeclareMathOperator{\Arg}{Arg}

\makeatother

\usepackage{babel}
\begin{document}

\title{WAVE ASYMPTOTICS AND THEIR APPLICATION TO ASTROPHYSICAL PLASMA LENSING}

\author{G. Grillo and J. M. Cordes}

\affil{Cornell Center for Astrophysics and Planetary Science and Department
of Astronomy, Cornell University, Ithaca, NY 14853, USA}
\begin{abstract}
Plasma lensing events can have significant observational consequences,
including flux density modulations and perturbations in pulse arrival
times. In this paper we develop and apply a formalism that extends
geometrical optics to describe the effects
of two dimensional plasma lenses of arbitrary shape. We apply insights
from catastrophe theory and the study of uniform asymptotic expansions
of integrals to describe the lensing amplification close to fold caustics
and in shadow regions, and explore the effects of image appearance
and disappearance at caustics in the time of arrival (TOA) perturbations due to lensing.
The enhanced geometric optics approach successfully reproduces the
predictions from wave optics and can be efficiently used to simulate
multifrequency TOA residuals during lensing events. Lensing will introduce
perturbations both in the way the residuals change as a function of
frequency and also in the magnitude and sign of the residuals averaged
over a frequency band. The deviations from the expected dispersive
$\nu^{-2}$ scaling will be most significant when including observations
at low frequencies. We examine the consequences of lensing in the
context of precision pulsar timing and touch on its potential relevance
to the study of FRBs.
\end{abstract}

\section{Introduction}\label{sec: 1}

The phenomenon of astrophysical plasma lensing has attracted considerable
attention ever since the first detections of so-called ``extreme
scattering events'' (ESEs) in the late 1980s (\citealt{Fiedleretal1987})
and early 1990s (\citealt{Cognardetal1993}), during which the measured
flux density of the observed objects (a millisecond pulsar in the
latter case, and a quasar in the former) underwent large fluctuations
with a frequency dependent structure over a period of time of the
order of months. Subsequent works describing observations of ESEs,
such as those by \citet{Fiedleretal1994} and \citet{Cleggetal1996}
mentioned the idea, introduced in \citet{Cognardetal1993}, that these
events were the result of plasma overdensities in the interstellar
medium that act as lenses as they cross the line of sight between
the Earth and the source of radiation, refracting the incoming radio
waves and creating observable regions of focusing and defocusing. 

\citet{Cleggetal1998} gave a detailed exposition of the geometric
optics of one dimensional Gaussian lenses and performed numerical
simulations to find appropriate lens parameters that could match the
observed flux fluctuations of specific ESEs, and some subsequent works
have also aimed to derive the characteristics of specific lenses deemed
to be responsible for particular ESE observations (\citealt{Pushkarevetal2013,Bannisteretal2016,Tuntsovetal2016,Vedanthametal2017,Kerretal2017,Mainetal2018}).

More recently, plasma lensing has also been suggested as a possible
mechanism to explain certain properties of FRBs (\citealt{Cordesetal2017,Dai&Lu2017}),
and other works have examined different kinds of lens models, as well
as their possible observational signatures (\citealt{Pen&King2012,Er&Rogers2017}),
although most of the analysis so far has been done in only one dimension
and for a few specific lens shapes.

Plasma lensing events do not only have observable effects in the source's
light curve, they also introduce perturbations in the times of arrival (TOAs)
of the radiation, via a combination of geometric and dispersive effects.
Thus, plasma lensing events can have potentially important consequences
for pulsar timing, as the possible detection of low frequency gravitational
waves via this method is dependent on our ability to detect $\lesssim100$
ns deviations in pulse arrival times. In fact, some plasma lensing
events have been inferred by their effects on observed pulsar TOAs
(\citealt{Lametal2018}) and dispersion measures (DMs) (\citealt{Colesetal2015}),
instead of their effects on measured flux density, since in some cases
the presence of strong scintillations can effectively mask whatever
effects the lensing events have on the source's light curve.

In contrast to the random fluctuations in the electron column density
that are responsible for scintillation, plasma lensing events are
produced by larger scale  inhomogeneities in the ISM, motivating
the use of geometrical optics. Nevertheless, it has been useful for
some authors modelling scintillation phenomena to study the effect
of nonturbulent phase screens, particularly in the transition regime
from weak to strong scintillations (\citealt{Watson&Melrose2006,Melrose&Watson2006}).
Furthermore, the underlying optics based on the Kirchhoff diffraction
integral (KDI) is the same for both scintillations and plasma lensing,
meaning that a considerable amount of the formalism used in the study
of scintillations can be applied in the latter context.

A potentially important effect of plasma lensing is the appearance
and disappearance of multiple images of the source as the lens crosses
the line of sight. Such multiple imaging has been directly observed
in cases in which the angular separation of some of the images has
been large enough (\citealt{Guptaetal1999,Pushkarevetal2013}), and
can be inferred from the existence of fringes in the dynamic spectra
of pulsars during certain epochs of observation (\citealt{Cordes&Wolszcan1986,Guptaetal1994,Cordesetal2006}).
The coalescence of images is associated with regions in which a straightforward
calculation of the flux using geometric optics diverges; these regions
are known as caustics, and the ability to describe these regions is
of importance both in the context of plasma lensing and scintillation
(\citealt{Goodmanetal1987,Melrose&Watson2006,Cordesetal2017}). The
geometric optics framework, however, is useful because it provides
information about the different images, including their amplitudes,
phases, and locations, and at the same time provides a relatively
simple way of calculating the total flux without the need of finding
a full solution to the KDI. Thus, it is desirable to describe the
amplification in the caustic regions without having to abandon the
geometric optics point of view. Different authors in the astrophysical
context have employed a variety of methods to handle the geometrical
optics infinities, but so far the problem has not been solved using
wave asymptotic methods derived from the geometrical theory of diffraction
(\citealt{Borovikov&Kinber1994}) and catastrophe optics (\citealt{Poston&Stewart1978,Berry&Upstill1980,Stamnes1986,Kravstov&Orlov1999,Katsaounisetal2001,Kryukovskiietal2006}),
in order
to predict the potential observational signatures
of two dimensional plasma lenses of arbitrary shape.

Our primary goal in this paper is therefore to use wave asymptotic
methods to characterize the effects of astrophysical plasma lensing,
develop the resulting formalism that describes the observational effects
of two dimensional plasma lenses that cross our line of sight, and
present some numerical results based on the application of this formalism.
We restrict ourselves to cases in which the source of radiation can
be accurately regarded as a point source, and focus on the effects
of plasma lensing on pulsar timing. The paper is divided as follows.
In \S\ref{sec: 2}, we present what we call the zeroth and first order
geometrical optics of two dimensional lenses, which formally yields
infinite flux amplitudes at caustic regions. In \S\ref{sec: 3} we
use wave asymptotic methods to construct a second order geometric
optics description. In \S\ref{sec: 4} we use the concepts developed
in \S\ref{sec: 2} and \S\ref{sec: 3} to examine the TOA and DM perturbations
due to a specific plasma lens realization, and we summarize conclusions
in \S\ref{sec: 5}. We expect to apply the methodology presented here
to specific events in subsequent work.

\section{Zeroth and first order geometric optics}\label{sec: 2}

\begin{figure}[t]
\begin{centering}
\includegraphics[width=0.9\columnwidth,height=0.9\columnwidth]{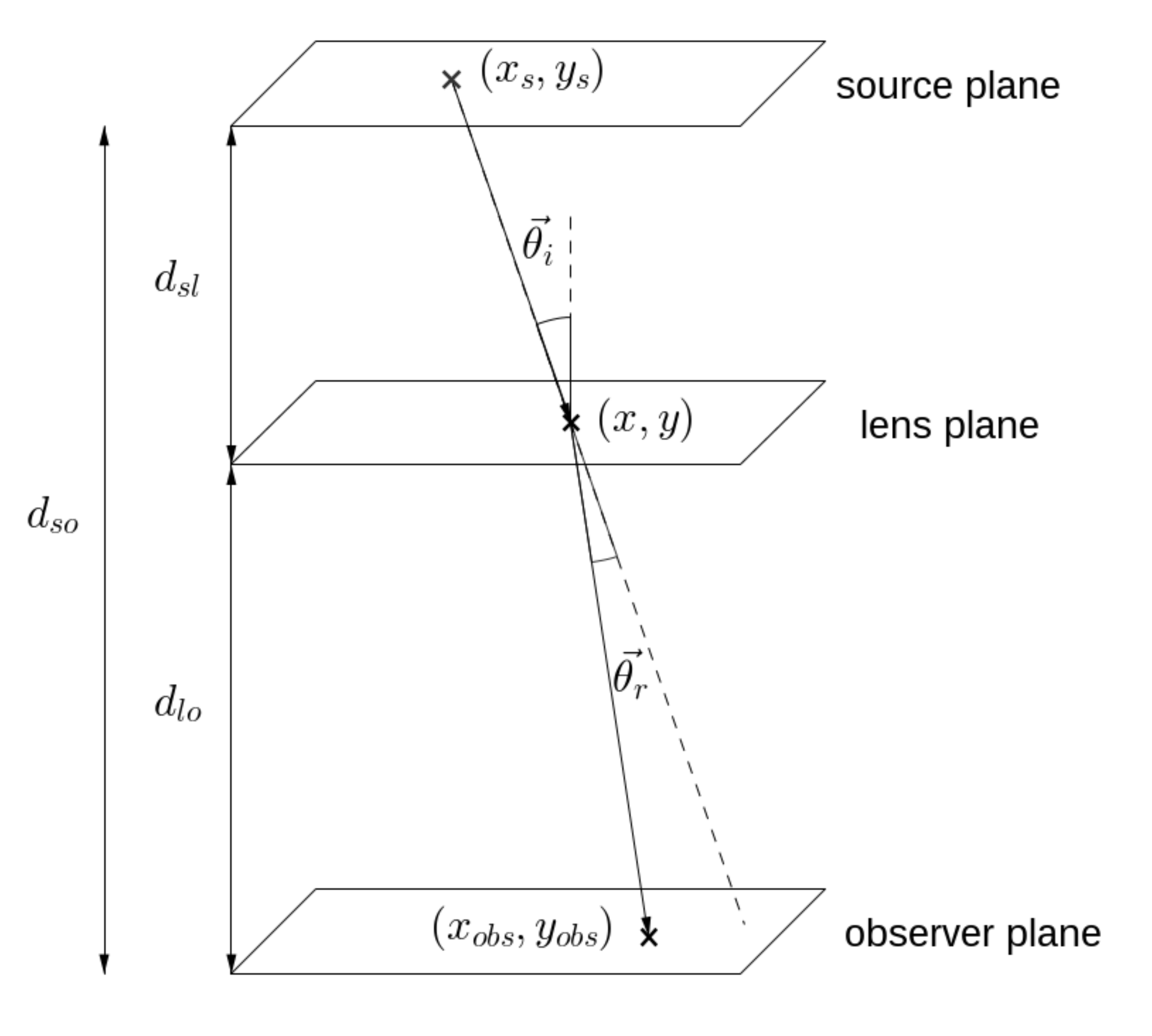}
\par\end{centering}
\caption{\label{fig:1}Lensing geometry.}
\end{figure}

\subsection{Geometrical picture}\label{subsec:2.1}

We follow the basics of the treatment given in \citealt{Cleggetal1998}
and \citealt{Cordesetal2017} but extend their results to two dimensions\footnote{\citet{Tuntsovetal2016} also gives a two dimensional account.}.
We start by defining planes for the source, the lens, and the observer
with coordinates $\boldsymbol{x_{s}}$, $\boldsymbol{x}$, and $\boldsymbol{x_{obs}}$,
respectively, with a source-lens distance $d_{sl}$, a lens-observer
distance $d_{lo}$, and a source-observer distance $d_{so}=d_{sl}+d_{lo}$,
as depicted in Figure \ref{fig:1}. The geometric optics approximation
treats the radiation emitted from the source as a cone of rays, and
the effects of lensing can be described by the way the lens affects
the mapping of the rays from the source plane to the observer plane.
From the geometry in the figure, we see that the 2D angle of incidence
of a ray into the lens plane $\boldsymbol{\theta_{i}}$ and  its
deviation angle are given (in the paraxial approximation) by

\begin{eqnarray}
\boldsymbol{\theta_{i}} & = & \frac{\boldsymbol{x_{s}}-\boldsymbol{x}}{d_{sl}}\label{eq: 1}\\
\boldsymbol{\theta_{r}} & = & \frac{\boldsymbol{x_{obs}}-\boldsymbol{x}}{d_{lo}}-\boldsymbol{\theta_{i}}\label{eq: 2}.
\end{eqnarray}

Combining into a single equation in terms of $\boldsymbol{\theta_{r}}$
gives the lens equation,
\begin{equation}
\boldsymbol{x_{s}}\left(\frac{d_{lo}}{d_{so}}\right)+\boldsymbol{x_{obs}}\left(\frac{d_{sl}}{d_{so}}\right)=\boldsymbol{x}+\boldsymbol{\theta_{r}}\left(\frac{d_{sl}d_{lo}}{d_{so}}\right)\label{eq: 3}.
\end{equation}

We now define a new set of coordinates $\boldsymbol{x'}$ as a combination
of the source and observer coordinates scaled by the distances, namely
\begin{equation}
\boldsymbol{x'}\equiv\boldsymbol{x_{s}}\left(\frac{d_{lo}}{d_{so}}\right)+\boldsymbol{x_{obs}}\left(\frac{d_{sl}}{d_{so}}\right)\label{eq: 4} .
\end{equation}
and write the lens equation in the simpler form

\begin{equation}
\boldsymbol{x'}=\boldsymbol{x}+\boldsymbol{\theta_{r}}\left(\frac{d_{sl}d_{lo}}{d_{so}}\right)\label{eq: 5},
\end{equation}

This expression is perfectly general and not only applies to plasma
lensing, but to gravitational lensing as well (\citealt{Schneideretal1992}).
The nature of the lensing is what determines the formula for the deviation
angle $\boldsymbol{\theta_{r}}$. A general expression for this angle
can be obtained with the additional assumptions that the lens's surface
slope is small, and that the lens's medium is uniform. The result
of the ray propagating through the lens is that the lens advances
or retards the ray's phase, depending on whether the value of the
refractive index $n_{r}$ is greater or smaller than unity, because
the phase velocity $v_{p}$ will be smaller or greater than $c$.
More precisely, we can write this phase difference $\delta\phi_{{\rm lens}}$
as
\begin{equation}
\delta\phi_{{\rm lens}}=\omega\tau=kc\tau\label{eq: 6} ,
\end{equation}
where $\tau$ is the 
propagation time difference between a
lensed ray and an unlensed ray, $k=2\pi/\lambda$ is the wavenumber
and $\omega$ is the radiation's angular frequency. By this definition,
$\tau<0$ implies that $\delta\phi_{{\rm lens}}<0$ and therefore
$v_{p}>c$. For a lens of length $l$ parallel to the direction of
propagation, this is
\begin{equation}
\tau=\frac{l}{c}\left(n_{r}-1\right)\label{eq: 7} .
\end{equation}

For a cold, unmagnetized plasma, the frequency dependent index of
refraction is given by
\begin{eqnarray}
n_{r} & = & \sqrt{1-\left(\frac{\omega_{e}}{\omega}\right)^{2}}\approx1-\frac{\lambda^{2}r_{e}n_{e}}{\pi}\label{eq: 8},
\end{eqnarray}
where $\omega_{e}^{2}=4\pi n_{e}e^{2}/m_{e}$ corresponds to the square
of the electron plasma frequency, $e$ is the elementary charge, $m_{e}$
is the mass of the electron, $r_{e}$ is the electron's classical
radius, and $n_{e}$ is the electron number density, and the approximate
equality comes from the fact that $\omega_{e}\ll\omega$ for $\omega$
within the radio spectrum. According to geometrical optics, rays propagate
in the direction normal to the surfaces of constant phase (\citealt[ Ch. 3]{Born&Wolf1999}),
so the refractive angle $\boldsymbol{\theta_{r}}$ is given by
\begin{equation}
\boldsymbol{\theta_{r}}=\frac{1}{k}\nabla\delta\phi_{{\rm lens}}\label{eq: 9} .
\end{equation}

When the electron column density or dispersion measure perturbation
${\rm DM}=n_{e}l$ at the lens plane varies as a function of transverse
position, ${\rm DM}\rightarrow{\rm DM}(\boldsymbol{x})$, $\boldsymbol{\theta_{r}}\neq0$,
and lensing occurs. Putting everything together, the phase perturbation
becomes
\begin{equation}
\delta\phi_{{\rm lens}}(\boldsymbol{x})=-\lambda r_{e}{\rm DM}(\boldsymbol{x})\label{eq: 10} ,
\end{equation}
which implies that the refractive angle is
\begin{align}
\boldsymbol{\theta_{r}} & =-\frac{\lambda^{2}r_{e}}{2\pi}\nabla{\rm DM}(\boldsymbol{x})=-\frac{c^{2}r_{e}}{2\pi\nu^{2}}\nabla{\rm DM}(\boldsymbol{x})\label{eq: 11} .
\end{align}

For convenience, we write ${\rm DM}(\boldsymbol{x})$ as the product
of a maximum perturbation ${\rm DM}_{\ell}$ and a function with unit
maximum $\psi(\boldsymbol{x})$, and take the origin of the lens plane's
coordinate system to coincide with the lens's center. Thus Eq. \ref{eq: 11}
takes the form
\begin{equation}
\boldsymbol{\theta_{r}}=-\frac{c^{2}r_{e}{\rm DM}_{\ell}}{2\pi\nu^{2}}\nabla\psi(\boldsymbol{x})\label{eq: 12} .\end{equation}

We now define the Fresnel scale as $r_{F}=\sqrt{cd_{sl}d_{lo}/2\pi d_{so}\nu}$,
the lens phase as $\phi_{0}=-cr_{e}{\rm DM}_{\ell}/\nu$, and a new
parameter $A=r_{F}^{2}\phi_{0}$, and substitute Eq. \ref{eq: 12}
in terms of these new quantities into the lens equation, which yields
a more compact form that is specific to plasma lensing, 
\begin{equation}
\boldsymbol{x'}=\boldsymbol{x}+A\nabla\psi(\boldsymbol{x})\label{eq: 13} .
\end{equation}

Finally, we define dimensionless coordinates using the characteristic
lens scales $a_{x}$ and $a_{y}$, such that $u_{x}'=x/a_{x}$ and
$u_{y}'=y/a_{y}$, and explicitly write Eq. \ref{eq: 13} in its adimensionalized
component form. Using the notation $\psi_{ij}=\nicefrac{\partial^{i+j}\psi}{\partial u_{x}^{i}\partial u_{y}^{j}}$,
and defining $\alpha_{x,y}=A/a_{x,y}^{2}$,
\begin{align}
\left[\begin{array}{c}
u_{x}'\\
u_{y}'
\end{array}\right] & =\left[\begin{array}{c}
u_{x}+\frac{A}{a_{x}^{2}}\psi_{10}(u_{x},u_{y})\\
u_{y}+\frac{A}{a_{y}^{2}}\psi_{01}(u_{x},u_{y})
\end{array}\right]\nonumber \\
 & =\left[\begin{array}{c}
u_{x}+\alpha_{x}\psi_{10}\\
u_{y}+\alpha_{y}\psi_{01}
\end{array}\right]\label{eq: 14} .
\end{align}

In general, Eq. \ref{eq: 14} must be solved numerically using a root
finding algorithm. More details on the numerical techniques used to
produce the examples presented throughout the paper can be found in Appendix
\ref{sec: Appendix C}. The vector $\boldsymbol{u'}(t)$ changes as
a function of time as the Earth, the lens, and the source move with
different velocities, and the nature of this change will partly determine
the observational signature of a specific lens realization. The number
of solutions of the equation corresponds to the number of images of
the source as seen by the observer, and in general vary as a function
of $\boldsymbol{u'}(t)$ and the parameters $\alpha_{x,y}$.

\subsection{Zeroth order gain}\label{subsec: 2.2}

A large majority of the existing literature on plasma lensing (\citealt{Cleggetal1998,Pen&King2012,Tuntsovetal2016,Cordesetal2017,Er&Rogers2017,Vedanthametal2017})
derives the gain (or magnification) for an individual image $G_{j}$
directly from some version of Eq. \ref{eq: 14}, and the total gain
is found by adding together the gains of all $n$ images. More specifically,
the image magnification is said to correspond to the absolute value
of the inverse of the Jacobian of the mapping between the $u$ and
$u'$ planes, evaluated at a solution to the lens equation $\boldsymbol{u}=\boldsymbol{u_{j}^{0}}$,
\begin{align}
G_{j} & =\left|\mathcal{J}\right|^{-1}\nonumber \\
 & =\left|\left(1+\alpha_{x}\psi_{20}\right)\left(1+\alpha_{y}\psi_{02}\right)-\alpha_{x}\alpha_{y}\psi_{11}^{2}\right|{}^{-1}\label{eq: 15}
\end{align}
and the total gain is
\begin{equation}
G=\sum_{j=1}^{n}G_{j}\label{eq: 16} .
\end{equation}

We refer to this expression as the ``zeroth order'' geometrical
optics gain. It corresponds to a sum of intensities, and as such it
fails to take into account the interference between the images that
arises from the phase differences of the corresponding fields. An
accurate description of the interference pattern can be obtained by
solving the Kirchhoff diffraction integral (KDI), which we introduce
below.

\subsection{The 2D Kirchhoff diffraction integral}\label{subsec: 2.3}

Once we adopt a wave description of the radiation, the scalar wavefield
as a function of position with respect to the source is given by the
time independent Helmholtz equation. The general form of the KDI is
a formal solution to the Helmholtz equation (\citealt[Ch. 8]{Born&Wolf1999};
\citealt[Ch. 8]{Thorne&Blandford2017}). In the paraxial approximation
and for the near field, as is the case for AU sized lenses and astronomical
distances, the integral can be written in terms of dimensionless coordinates
(\citealt{Goodmanetal1987,Melrose&Watson2006,Cordesetal2017})
\begin{equation}
\varepsilon(\boldsymbol{u'},\nu)=\frac{a_{x}a_{y}}{2\pi r_{F}^{2}}\iint d^{2}u\exp(i\Phi)\label{eq: 17} ,
\end{equation}
where the phase $\Phi$ is the sum of a geometric term and the phase
perturbation due to the lens, $\delta\phi_{{\rm lens}}=\phi_{o}\psi(\boldsymbol{u})$,
\begin{equation}
\Phi(\boldsymbol{u'},\boldsymbol{u},\nu)=\frac{1}{2r_{F}^{2}}\left[a_{x}^{2}(u_{x}-u_{x}')^{2}+a_{y}^{2}(u_{y}-u_{y}')^{2}\right]+\phi_{0}\psi(\boldsymbol{u})\label{eq: 18} .
\end{equation}

The integral is normalized such that in the absence of a lens (ie.
$\delta\phi_{{\rm lens}}=0$), $\varepsilon(\boldsymbol{u'},\nu)=1$
for all $\boldsymbol{u'}$ and $\nu$. Analytic solutions to the integral
are only available for a few specific forms of $\psi$ (\citealt{Watson&Melrose2006}).
As detailed in Appendix \ref{sec: Appendix A}, this representation of $\Phi$
allows us to write the integral as a convolution of two functions,
which can then be solved numerically by employing the convolution
theorem and the Fast Fourier Transform (FFT). However, this method
is only adequate for lenses that have sizes that are a small fraction
of an AU and in cases where $\left|\phi_{0}\right|$ is relatively
small, because the required grid size for proper sampling grows prohibitively
large as the oscillations of $\exp(i\Phi)$ become more pronounced.

An approximate solution that grows more accurate as the strength of
the lens increases follows by applying the method of stationary phase.
For a rapidly oscillating two dimensional integral of the form $I(\boldsymbol{x})=\iint d^{2}\boldsymbol{x}g(\boldsymbol{x})\exp[if(\boldsymbol{x})]$,
the stationary phase lemma (\citealt{Bleistein&Handelsman1975}) indicates
that the principal contributions to the integral's value come from
the points in which the phase is stationary, that is, where the derivatives
of the phase vanish, $f_{10}=f_{01}=0$. In the general case where
these points $\boldsymbol{x}=\boldsymbol{x_{j}^{0}}$ are complex,
each provides a contribution to the integral $I_{j}$ given by (\citealt{Connor1973a})
\begin{equation}
I_{j}=\frac{2\pi ig_{j}\exp\left(if_{j}\right)}{\Delta_{j}^{\nicefrac{1}{2}}}\label{eq: 19},
\end{equation}
where $f_{j}=f(\boldsymbol{x_{j}^{0}})$, $g_{j}=g(\boldsymbol{x_{j}^{0}})$,
$\Delta_{j}=f_{20}f_{02}-f_{11}^{2}$ evaluated at $\boldsymbol{x_{j}^{0}}$,
and the square root in the denominator is taken to be positive or
negative depending on the context. When the stationary point is purely
real, the contribution reduces to (\citealt{Bleistein&Handelsman1975,Cooke1982})
\begin{equation}
I_{j}=\frac{2\pi g_{j}\exp\left[if_{j}+\frac{i\pi}{4}(\delta_{j}+1)\sigma_{j}\right]}{\left|\Delta_{j}\right|^{\nicefrac{1}{2}}}\label{eq: 20},
\end{equation}
where $\sigma_{j}=\sgn(\Delta_{j})$, $\delta_{j}=\sgn(f_{02})$,
and the square root in the denominator is now taken to be positive.
In the case of the KDI as given in the form Eq. \ref{eq: 17}, the
points of stationary phase correspond to the points that satisfy the
two dimensional equation
\begin{equation}
\left[\begin{array}{c}
\Phi_{10}\\
\Phi_{01}
\end{array}\right]=\left[\begin{array}{c}
{\displaystyle \frac{a_{x}^{2}(u_{x}-u_{x}')}{r_{F}^{2}}}+\phi_{0}\psi_{10}\\
{\displaystyle \frac{a_{y}^{2}(u_{y}-u_{y}')}{r_{F}^{2}}}+\phi_{0}\psi_{01}
\end{array}\right]=\left[\begin{array}{c}
0\\
0
\end{array}\right]\label{eq: 21} .
\end{equation}

\begin{figure*}
\begin{centering}
\includegraphics[width=0.9\textwidth]{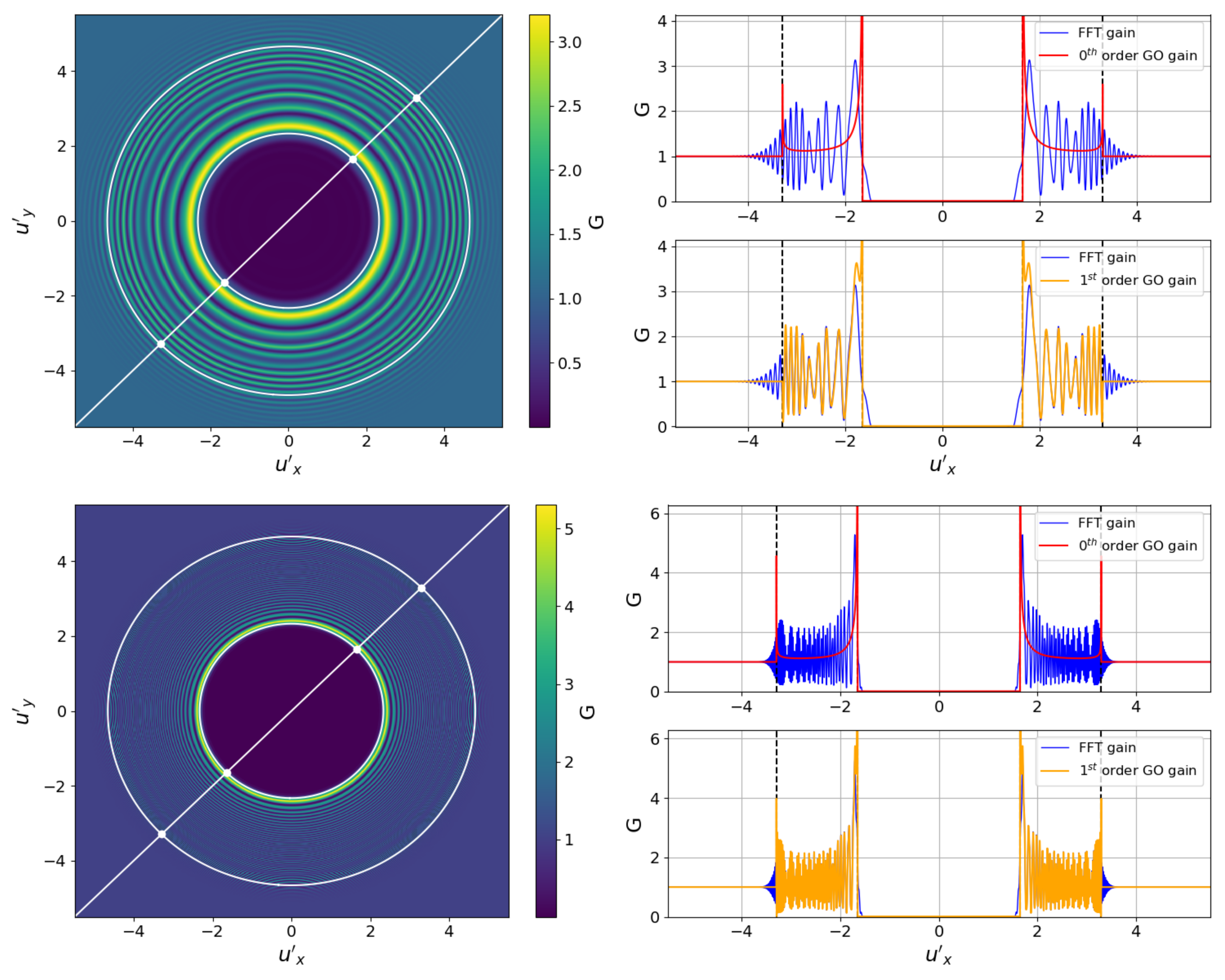}
\par\end{centering}
\caption{\label{fig:2}Comparison of the gains obtained from a full numerical
solution of the KDI, zeroth order geometrical optics, and first order
geometrical optics. The top panel corresponds to a lens with $\phi_{0}=-50$
rad and the bottom panel corresponds to one with $\phi_{0}=-250$
rad (thus ${\rm DM}_{\ell}>0$ in both cases, and the lenses are diverging).
The frequency of observation is $\nu=0.8$ GHz, $d_{so}=1$ kpc, $d_{sl}=0.5$
kpc for both the top and bottom panels. For the top panel, $a_{x}=a_{y}=1.5\times10^{-2}$
AU, and for the bottom panel, $a_{x}=a_{y}=1.5\sqrt{5}\times10^{-2}$
AU. The lens shape is described by a two dimensional Gaussian, $\psi(\boldsymbol{u})=\exp\left(-u_{x}^{2}-u_{y}^{2}\right)$.
The left column shows color maps of the gain obtained by solving the
KDI via the FFT. The white circles correspond to caustic curves, and
the straight white line shows the path of the observer through the
$u'$ plane. The right column shows the gain along this path as calculated
via the FFT method, zeroth order geometrical optics, and first order
geometrical optics. The points of intersection between the caustics
and the observer path are marked by white points in the left column
and by dashed black lines on the plots in the right column. The geometric
optics gain at the caustics is formally infinite, so the GO gains
were evaluated up to a short distance away from the caustic.}
\end{figure*}

A quick examination reveals that this is precisely equivalent to the
lens equation Eq. \ref{eq: 14}, given our definitions of the parameters
$\alpha_{x,y}$, which therefore implies that solving the KDI by the
method of stationary phase leads to geometric optics\footnote{It is also possible to derive the geometric optics quantities by directly
solving the Helmholtz equation via WKB methods (see, e.g. \citealt[Ch. 3]{Born&Wolf1999};
\citealt{Katsaounisetal2001}; \citealt[Ch. 12]{Poston&Stewart1978}).}. Performing the appropriate substitutions in Eq. \ref{eq: 21}, we
have that the scalar field $\varepsilon_{j}^{r}$ due to a real stationary
point is
\begin{align}
\varepsilon_{j}^{r}(\boldsymbol{u'},\nu) & =\frac{a_{x}a_{y}}{r_{F}^{2}\left|\Delta_{j}\right|^{\nicefrac{1}{2}}}\exp\left[i\Phi_{j}+\frac{i\pi}{4}(\delta_{j}+1)\sigma_{j}\right]\nonumber
\end{align}
\begin{align}
 & =\frac{\exp\left[i\Phi_{j}+\frac{i\pi}{4}(\delta_{j}+1)\sigma_{j}\right]}{\left|\left(1+\alpha_{x}\psi_{20}\right)\left(1+\alpha_{y}\psi_{02}\right)-\alpha_{x}\alpha_{y}\psi_{11}^{2}\right|^{\nicefrac{1}{2}}}\label{eq: 22},
\end{align}
where now we have $\Delta_{j}=\Phi_{20}\Phi_{02}-\Phi_{11}^{2}$,
$\sigma_{j}=\sgn(\Delta_{j})$, $\delta_{j}=\sgn(\Phi_{02})$, and
all quantities are evaluated at the stationary points, $\boldsymbol{u}=\boldsymbol{u_{j}^{0}}$.
This gives the normalized scalar field due to one image of the source,
with a maximum amplitude 
\begin{equation}
A_{j}=\left|\mathcal{J}\right|^{\nicefrac{-1}{2}}=\frac{a_{x}a_{y}}{r_{F}^{2}|\Delta_{j}|^{\nicefrac{1}{2}}}\label{eq: 23}
\end{equation}
and an oscillating component with phase 
\begin{equation}
\beta_{j}^{r}=\Phi_{j}+\frac{\pi}{4}(\delta_{j}+1)\sigma_{j}\label{eq: 24} .
\end{equation}

The total scalar field due to real solutions of the lens equation
is simply the sum of the contributions from the $n$ real stationary
points,
\begin{equation}
\varepsilon^{r}(\boldsymbol{u'},\nu)=\sum_{j=1}^{n}\varepsilon_{j}^{r}=\sum_{j=1}^{n}A_{j}e^{i\beta_{j}^{r}}\label{eq: 25} .
\end{equation}

The gain can then be obtained by taking the squared modulus of this
last expression, $G=\left|\varepsilon^{r}(\boldsymbol{u'},\nu)\right|^{2}$.
This is the ``first order'' geometrical optics gain. The presence
of the oscillatory component in each of the images results in interference.
As noted above, this is not correctly captured by Eq. \ref{eq: 16}.

\subsection{Accuracy and regions of applicability}\label{subsec: 2.4}

A curious feature of the phasors that emerge from the stationary phase
solutions is that they include not only the geometric phase $\Phi$
but also a potential phase shift related to the signs of the second
derivatives at the stationary points. This phase shift is physically
associated with the passage of a ray through a caustic. A caustic
corresponds to a surface in parameter space that yields a null Jacobian,
$\mathcal{J}=0$ (\citealt{Berry&Upstill1980,Kravstov&Orlov1999}).
As we approach a caustic, $A_{j}\rightarrow\infty$, and the zeroth
and first order geometric optics approximation fail. The reason for
this failure is that the approximations do not take into account diffractive
effects that occur due to the finite frequency of the waves. Caustics
also correspond to boundaries that separate regions in parameter space
that contain different numbers of real solutions to the lens equation.

Figure \ref{fig:2} illustrates the difference between the gains obtained
from the zeroth and first order approaches in the case of an overdense
(${\rm DM}_{\ell}>0)$, two dimensional Gaussian lens described by
$\psi(\boldsymbol{u})=\exp\left(-u_{x}^{2}-u_{y}^{2}\right)$ with
equal lens scales, $a_{x}=a_{y}$. For a fixed frequency of observation,
the wave optics amplification as a function of $\boldsymbol{u'}$
can be calculated by solving the KDI using the FFT. The left column
shows this amplification as a function of $\boldsymbol{u'}$ for the
range $-5.5\leq u_{x}'\leq5.5$ and two different values of $\phi_{o}$,
-50 rad (top) and -250 rad (bottom). The white circles correspond
to the caustics, and the straight white line denotes the observer's
path along the plane. The right column shows the zeroth (red) and
first (orange) order gains along the path superposed with the wave optics
gain (blue) for both cases.

From the figure, we can see that unlike the zeroth order approximation,
the first order approach is able to reproduce the wave optics oscillations
accurately in bright regions that contain more than one real solution
to the lens equation. However, wave optics also predicts that in regions
with only one real image of the source, the observer should still
see an interference pattern that decays (grows) exponentially as she
crosses from the caustic's bright (dark) side to the dark (bright)
side.

For instance, the Gaussian lens from the figure shows two sets of
circular caustics in the $u'$ plane. An observer crossing the $u'$
plane through its center will pass through three regions in which
the form of the gain is qualitatively different. Far away from the
two caustic zones, $G=1$, and the intensity shows no modulations
due to lensing. This corresponds to the dark side of the outer caustic.
As the observer approaches the outer caustic singularity, the intensity
starts showing oscillations whose amplitude grows exponentially, even
though there is still only one real solution to the lens equation.
Crossing into the region between the two caustics, the oscillations's
amplitude reach a peak shortly after the boundary, and the observer
sees three images corresponding to three real solutions to the lens
equation. After that, the amplitude decays and then recovers, peaking
right next to the boundary that separates the bright region from the
dark side of the inner caustic. This dark region contains a single,
highly demagnified image of the source, but there is still a hint
of an exponentially decaying interference pattern that disappears
a short distance away from the boundary. After crossing the center,
an equivalent pattern is observed in reverse as the observer moves
from the inner dark side to the bright region and then to the outer
dark side. \citet{Cleggetal1998}, \citet{Melrose&Watson2006}, and
\citet{Cordesetal2017} studied an analogous lens shape in one dimension.
The short paper by \citet{Stinebringetal2007} presents similar two
dimensional plots without the geometrical optics curves.

In summary, then, although the inclusion of ray interference dramatically
improves the accuracy of the geometric optics approximation, this
approach is still unable to reproduce the correct form of the gain
close to the caustic singularity (where it blows up), and on the dark
side of caustic boundaries (where it fails to account for oscillations).

\section{Second order geometric optics}\label{sec: 3}

\subsection{Complex rays}\label{subsec: 3.1}

So far, we have limited our analysis to the case in which coordinates
in the $u$ plane, and the solutions to the lens equation are purely
real. In order to reproduce the oscillations that occur in the caustic
shadows, however, it is necessary to extend the analysis to the complex
plane. When two or more real roots of the lens equation merge at a
caustic, they reemerge at the dark side as a complex conjugate pair
of solutions to the lens equation that yield a complex phase $\Phi_{\pm}=\Phi_{r}\pm i\Phi_{i}$.
$\Phi_{i}>0$ grows as we move farther into the shadow side in parameter
space. From Eq. \ref{eq: 19}, we can write the field $\varepsilon_{\pm}^{c}$
due to this complex conjugate pair as
\begin{align}
\varepsilon_{\pm}^{c}(\boldsymbol{u'},\nu) & =Ae^{\mp\Phi_{i}}e^{i\beta_{\pm}^{c}}\label{eq: 26},
\end{align}
where $A=a_{x}a_{y}\left|\Delta_{\pm}\right|^{-\nicefrac{1}{2}}/r_{F}^{2}$
is the same as in the case of a purely real stationary point (Eq.
\ref{eq: 24}), and $\beta_{\pm}^{c}=\Phi_{r}+\pi/2-\Arg\left(\Delta_{\pm}\right)/2$.
This expression implies that $\varepsilon_{+}^{c}$ decreases exponentially
as a function of $\Phi_{i}$, whereas $\varepsilon_{-}^{c}$ increases
exponentially. The exponentially increasing solution can be disregarded
as unphysical (\citealt{Kravstovetal1999}), but the exponentially decaying
contribution can be included as part of the asymptotic approximation
to the KDI. Doing so effectively reproduces the shadow side oscillatory
pattern predicted by wave optics, as long as we remain far enough
away from the caustic. At the caustic, the complex conjugate pair
of solutions merge, and $A\rightarrow\infty$.

The idea of looking for complex solutions to the lens equation has
surfaced in a variety of contexts. \citet{Schramm&Kayser1995} apply
the concept to gravitational lensing, and \citet{Budden&Terry1971}
apply it in the context of radio ray tracing in the atmosphere. There
is also a direct connection between complex stationary points, the
method of steepest descent, and hyperasymptotics of oscillatory integrals
(\citealt{Kaminsky1994,Howls1997}).

\subsection{Caustic location and extent of the caustic zone}\label{subsec: 3.2}

In the language of geometric optics, caustics correspond to envelopes
of families of rays, and are formed at the surfaces on which rays
cross each other. Determining the parameter values for ray crossings
to occur is, in general, a non trivial problem in more than one dimension
and for an arbitrary lens shape. For a fixed frequency of observation,
the necessary condition is that 
\begin{equation}
\left(1+\alpha_{x}\psi_{20}\right)\left(1+\alpha_{y}\psi_{02}\right)-\alpha_{x}\alpha_{y}\psi_{11}^{2}=0\label{eq: 27}
\end{equation}
for at least some value of $\boldsymbol{u}$. If this is the case,
caustic curves will show up in the $u$ plane, and their form in the
$u'$ plane can be determined by mapping these curves via the lens
equation. The caustic curves plotted over the colormaps in the left
column of Figure \ref{fig:2} were constructed using this method.

On the other hand, for a fixed $u'$ coordinate, the locations of
caustics in the frequency line need to be determined by solving the
set of equations
\begin{align}
\frac{\psi_{10}\psi_{01}}{\Delta u_{x}\Delta u_{y}}+\frac{\psi_{20}\psi_{01}}{\Delta u_{y}}+\frac{\psi_{02}\psi_{10}}{\Delta u_{x}}+\psi_{20}\psi_{02}-\psi_{11}^{2} & =0\nonumber \\
\left(\frac{a_{x}}{a_{y}}\right)^{2}\frac{\Delta u_{x}}{\psi_{10}}-\frac{\Delta u_{y}}{\psi_{01}} & =0\label{eq: 28}
\end{align}
for $\boldsymbol{u}$, where $\Delta u_{x,y}=u_{x,y}'-u_{x,y}$. The
caustics will be located at frequencies $\nu_{{\rm caus}}$, given
by
\begin{eqnarray}
\nu_{{\rm caus}} & = & \frac{c}{a_{x}}\left[\frac{d_{sl}d_{lo}r_{e}{\rm DM}_{\ell}}{2\pi d_{so}\Delta u_{x}}\psi_{10}\right]^{\nicefrac{1}{2}}\nonumber \\
 & = & \frac{c}{a_{y}}\left[\frac{d_{sl}d_{lo}r_{e}{\rm DM}_{\ell}}{2\pi d_{so}\Delta u_{y}}\psi_{01}\right]^{\nicefrac{1}{2}}\label{eq: 29},
\end{eqnarray}
evaluated at the solutions of Eq. \ref{eq: 28} for which the argument
under the square root is positive. Numerical results indicate that
the formation of caustics at fixed frequencies, for lenses with Gaussian-like
shapes (with a maximum electron column density at the center that
falls off relatively quickly) occurs when $\alpha_{x,y}\lesssim-1.2$
(for the positive ${\rm DM}_{\ell}$ case) and $\alpha_{x,y}\apprge0.5$
(for the negative ${\rm DM}_{\ell}$ case). If both $\alpha_{x}$
and $\alpha_{y}$ satisfy this condition, two sets of caustics form;
if only one does, just one set appears.

\begin{figure*}
\begin{centering}
\includegraphics[width=0.9\textwidth]{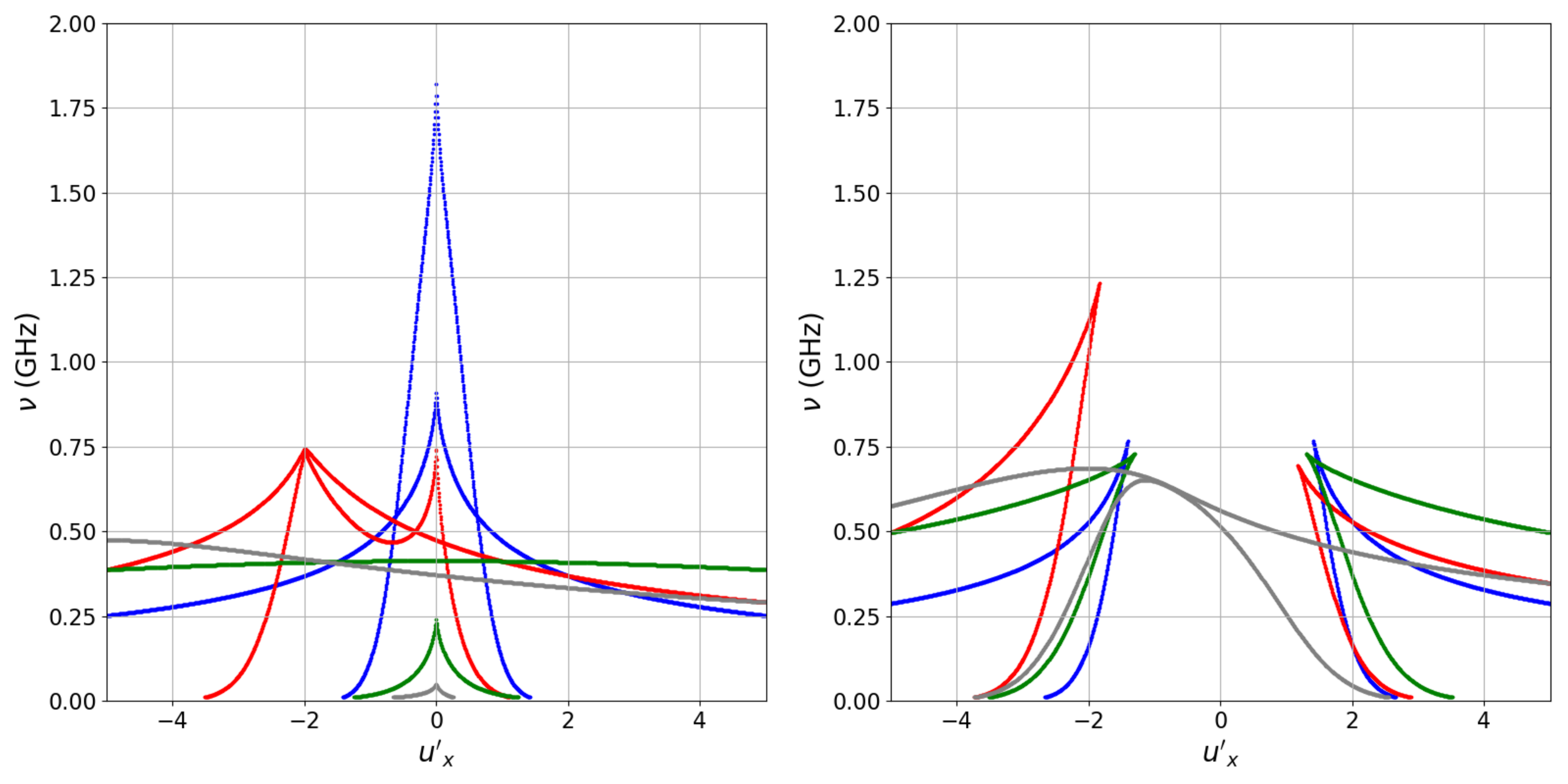}
\par\end{centering}
\caption{\label{fig: 3}Caustic curves in the dynamic spectra of underdense
(left) and overdense (right) two dimensional Gaussian lenses for different
paths across the $u'$ plane. The blue caustics derive from a path
with slope $m=1$ and y-intercept $n=0$, the red caustics have $m=0.5$
and $n=1$, the green caustics correspond to $m=0$, $n=1.5$, and
the grey caustics are produced by $m=0.3$ and $n=2$. We use a value
of ${\rm DM}_{\ell}=\pm10^{-3}$ pc cm\protect\textsuperscript{-3},
which corresponds to a lens phase $\phi_{0}\approx\mp3\times10^{4}$
rad. The source-observer distance $d_{so}=1$ kpc, and the source-lens
distance $d_{sl}=0.5$ kpc in both cases. Both lenses have scales
$a_{x}=0.5$ AU and $a_{y}=1$ AU.}
\end{figure*}

A consequence of this requirement is that larger lenses require larger
magnitudes of ${\rm DM}_{\ell}$ in order to form caustics in the
$u'$ plane at a fixed frequency of observation. Thus, small values
of ${\rm DM}_{\ell}$ will only lead to caustic formation in cases
involving small lenses or highly elongated lenses. For example, keeping
the relevant distances fixed at $d_{so}=1$ kpc and $d_{sl}=0.5$
kpc, a value of ${\rm DM}_{\ell}=\pm10^{-6}$ pc cm\textsuperscript{-3},
which corresponds to a lens phase of $\phi_{0}\approx\mp33$ rad at
$0.8$ GHz, yields a maximum value of $a_{x,y}\approx2.4\times10^{-2}$
AU for the overdense case and $a_{x,y}\approx3.6\times10^{-2}$ AU
for the underdense case. Ray crossings for lenses with $a_{x,y}\approx1$
AU would require a minimum value of $\left|{\rm DM}_{\ell}\right|\approx2\times10^{-3}$
pc cm\textsuperscript{-3} for the diverging lens and $\left|{\rm DM}_{\ell}\right|\approx7\times10^{-4}$
pc cm\textsuperscript{-3}for the converging lens, which correspond
to lens phases at $0.8$ GHz of $\phi_{o}\approx-5.8\times10^{4}$
rad and $\phi_{o}\approx2.4\times10^{4}$ rad, respectively. Changing
the lens-observer distance $d_{so}$ and source-lens distance $d_{sl}$
also leads to changes in $\alpha_{x,y}$, although not in a very simple
way because the value of $d_{lo}=d_{so}-d_{sl}$ also factors into
the expression. In general, however, if we keep $d_{sl}$ fixed at
$d_{so}/2$, increasing $d_{so}$ also increases $\alpha_{x,y}$ and
makes caustic formation more likely. The radius of the caustic curves
tends to increase linearly with $\left|\alpha_{x,y}\right|$\footnote{An important exception is the underdense (${\rm DM}_{\ell}<0)$ Gaussian
lens with $a_{x}=a_{y}$, which presents an infinitely small caustic
at the center corresponding to a focus, and a single circular caustic
surrounding it.}.

Figure \ref{fig: 3} shows the caustics in the dynamic spectra of
a lensing event for underdense (left) and overdense (right) Gaussian
lenses for multiple paths along the $u'$ plane, constructed by repeated
application of Eq. \ref{eq: 28} and Eq. \ref{eq: 29} over a range
of $u'$ coordinates. Although the lens parameters are identical in
both cases, it can be seen that flipping the sign of ${\rm DM}_{\ell}$
generates a completely different set of caustic curves, and that the
path of the observer through the $u'$ plane can also significantly
alter the caustic shapes.

Caustics will show up as a function of $\nu$ at a fixed value of
$u'$ if we search within a range of frequencies that contains a value
of $\nu$ that leads to at least one of the $\alpha_{x,y}$ parameters
having  a magnitude larger than the required minimum. Since $\left|\alpha_{x,y}\right|\propto\nu^{-2}$,
caustic curves in dynamic spectra, such as the ones depicted in Figure
\ref{fig: 3}, will  show up only at low frequencies. 

In practical terms, it is useful to be able to locate caustics as
functions of both $\boldsymbol{u'}$ and $\nu$. Telescope observations
made during an observing epoch correspond roughly to observations
made at a fixed $\boldsymbol{u'}$. Observations with a large enough
frequency range would in principle allow us to see the effects
of caustics (under the right circumstances) in a single epoch of observation
if a lensing event is taking place. At the same time, since the coordinates
in $u'$ change as a function of time, we also expect to see caustic
effects in observations made within a narrow frequency band over a
range of epochs.

At a caustic boundary, two or more images of the source appear or
merge, depending on whether the caustic is crossed from one side or
the other. In other words, the number of real roots of the lens equation
changes by at least two. The first order geometric approximation breaks
down in the vicinity of the caustic when two or more images of the
source become indistinguishable from each other. As noted by \citet{Kravstov&Orlov1999},
a useful operational definition for the width of the caustic zone
is the boundary at which the absolute value of the geometrical phase
difference $\left|\Delta\Phi_{ij}\right|$ between two or more roots
is less than $\pi$,
\begin{equation}
\left|\Delta\Phi_{ij}\right|\lesssim\pi\label{eq: 30},
\end{equation}
where $i,j$ are the labels of each of the roots. The number of coalescing
images determines the type of caustic, as it describes the kind of
singularity, or catastrophe, that occurs within the caustic zone.

A number of previous works (\citealt{Chako1965,Bleistein&Handelsman1975,Cooke1982,Wong2001,Cordesetal2017})
have dealt with the problem of obtaining the maximum gain within this
region by employing an extension of the stationary phase method to
approximate the gain at the singularity. Although the derived formulae
(some of which are presented in Appendix \ref{sec: Appendix B}) are relatively
simple to apply and can be useful for some types of analyses, it is
not in general correct because the geometric optics approximation
breaks down some distance away from the caustic, close to the boundary
defined by Eq. \ref{eq: 30}.

\subsection{Gain inside the caustic zone: catastrophe theory and uniform asymptotics}\label{subsec: 3.3}

Catastrophe theory, first developed by the mathematician Ren\'e Thom
(\citealt{Thom1972}) and subsequently applied to optics by Sir Michael
Berry and others (\citealt{Berry1976,Nye1978,Berry&Upstill1980}),
provides a useful way of categorizing geometric optics singularities.
The basic idea is that close to a caustic, the phase function can
be locally mapped into a standard form that is determined by the number
of merging images. This standard form is expressed in terms of a fixed
number of state and control variables, which are related by the mapping
to the physical variables. Solving the KDI for the particular case
of this standard form yields a transitional approximation that describes
the gain within the caustic region.

In general, it is very difficult to rigourously construct a mapping
that takes the global form of the phase to the standard form. Instead,
the mapping is performed by expanding the phase in a Taylor series
at the point that satisfies both the lens equation and Eq. \ref{eq: 27},
in addition to rotating and scaling the coordinate system such that
it is possible to match the coefficients present in this form of the
phase to the standard form of the catastrophe. This procedure is described
in \citet{Kravstov&Orlov1999}, and performed specifically for the
case of two dimensional scattering screens in the context of scintillation
by \citet{Goodmanetal1987}.

\citet{Watson&Melrose2006} rely on an analogous procedure to derive
the one dimensional transitional approximation for the case of two
merging images, which corresponds to a fold caustic. The fold catastrophe
is the first of the seven elementary catastrophes described by Thom
in his original work, and it is the simplest to model and describe.
In the vicinity of the fold, the phase can be locally mapped to a
cubic, and the KDI can be mapped into the canonical integral (\citealt{Berry&Upstill1980})
\begin{align}
I_{{\rm fold}}(\xi) & =\frac{1}{\sqrt{2\pi}}\int_{-\infty}^{\infty}dt\exp\left[i\left(\frac{t^{3}}{3}+\xi t\right)\right]\nonumber \\
 & =\sqrt{2\pi}\Ai(\xi)\label{eq: 31},
\end{align}
where $\xi$ denotes a control variable and $t$ denotes a state variable,
and $\Ai(\xi)$ is the Airy function. The observer sees no real images
on the dark side of the caustic, and two images on the bright side,
but the intensity at the dark side does not drop to zero instantly
as predicted by first order geometric optics. For practical purposes,
it is possible to adopt this transitional form within the caustic
region, and revert back to the regular geometrical optics description
far away from the caustic, as in \citet{Watson&Melrose2006}.

A better, more general solution is to employ the method of uniform
asymptotics, as initially developed by \citet{Chester1957}, \citet{Ursell1965},
and \citet{Ludwig1966} for oscillatory integrals, and later explicitly
applied to optics and related to catastrophe theory by \citet{Kravstov1968} and \citet{Kravstov&Orlov1999}.
This solution enables us to describe the gain in regions both close
and far away from the caustics by the application of a single, global
expression that employs the integral of the standard form associated
with the type of catastrophe involved, the derivatives of this integral,
and some combination of the parameters derived from geometrical optics.
Close to the caustic, the expression behaves like the transitional
approximation, and far away from it, it matches the field given by
the regular geometrical optics approximation. 

Uniform asymptotic expressions for the fold caustic have been derived
by multiple authors starting with \citet{Chester1957}\footnote{Also see \citet{Ludwig1966,Connor1973a,Stamnes1986,Borovikov&Kinber1994,Kravstov&Orlov1999,Qiu&Wong2000,Katsaounisetal2001} },
and in general there are slight variations between each of the presented
expressions. We derive it here in an intuitive manner.

The general scheme consists in starting with an ansatz with the same
number of terms as there are rays involved in the formation of the
caustic, one term involving the function corresponding to the canonical
caustic integral, and the rest involving its derivatives. Each of
these terms is multiplied by an unknown coefficient, and their sum
is multiplied by a phasor. For the fold caustic, it is possible to
construct the uniform asymptotic simply by starting with the ansatz
and matching the relevant parameters to the geometrical optics coefficients
far away from the caustic, by employing the asymptotic forms of the
Airy function and its derivative for large negative and positive arguments.
Thus, for the bright side, we start with an ansatz of the form,
\begin{align}
\varepsilon_{{\rm bright}}(\boldsymbol{u'},\nu) & =e^{i\chi}\left[g_{1}I_{{\rm fold}}(\xi)+g_{2}I_{{\rm fold}}'(\xi)\right]\nonumber \\
 & =\sqrt{2\pi}e^{i\chi}\left[g_{1}\Ai(\xi)+g_{2}\Ai'(\xi)\right]\label{eq: 32},
\end{align}
where $g_{j}$, $\chi$, and $\xi$ are all potentially functions
of $\boldsymbol{u'}$. From Eq. \ref{eq: 25}, we have that the first
order geometrical optics solution in the case of two real rays can
be written as
\begin{align}
\varepsilon^{r}(\boldsymbol{u'},\nu) & =A_{1}e^{i\beta_{1}^{r}}+A_{2}e^{i\beta_{2}^{r}}\label{eq: 33} .
\end{align}

The asymptotic forms of the Airy function and its derivative for large
negative argument are the well known formulas,
\begin{align}
\Ai(\xi) & \approx\frac{1}{\sqrt{\pi}}(-\xi)^{\nicefrac{-1}{4}}\cos\left[\frac{2}{3}(-\xi)^{\nicefrac{3}{2}}-\frac{\pi}{4}\right]\label{eq: 34}\\
\Ai'(\xi) & \approx\frac{1}{\sqrt{\pi}}(-\xi)^{\nicefrac{1}{4}}\sin\left[\frac{2}{3}(-\xi)^{\nicefrac{3}{2}}-\frac{\pi}{4}\right]\label{eq: 35},
\end{align}
which are obtained by applying the one dimensional stationary phase
method to the integral in Eq. \ref{eq: 31}. Defining $\gamma=2(-\xi)^{\nicefrac{3}{2}}/3-\pi/4$,
using Euler's identity,  and substituting into Eq. \ref{eq: 32}, we
get
\begin{eqnarray}
\varepsilon_{{\rm bright}}(\boldsymbol{u'},\nu) & = & \frac{e^{i\chi}}{\sqrt{2}}\left\{ e^{i\gamma}\left[g_{1}(-\xi)^{\nicefrac{-1}{4}}+ig_{2}(-\xi)^{\nicefrac{1}{4}}\right]\right.\nonumber \\
 & = & \left.+e^{-i\gamma}\left[g_{1}(-\xi)^{\nicefrac{-1}{4}}-ig_{2}(-\xi)^{\nicefrac{1}{4}}\right]\right\} \label{eq: 36} .
\end{eqnarray}

\begin{figure*}[t]
\begin{centering}
\includegraphics[width=0.9\textwidth]{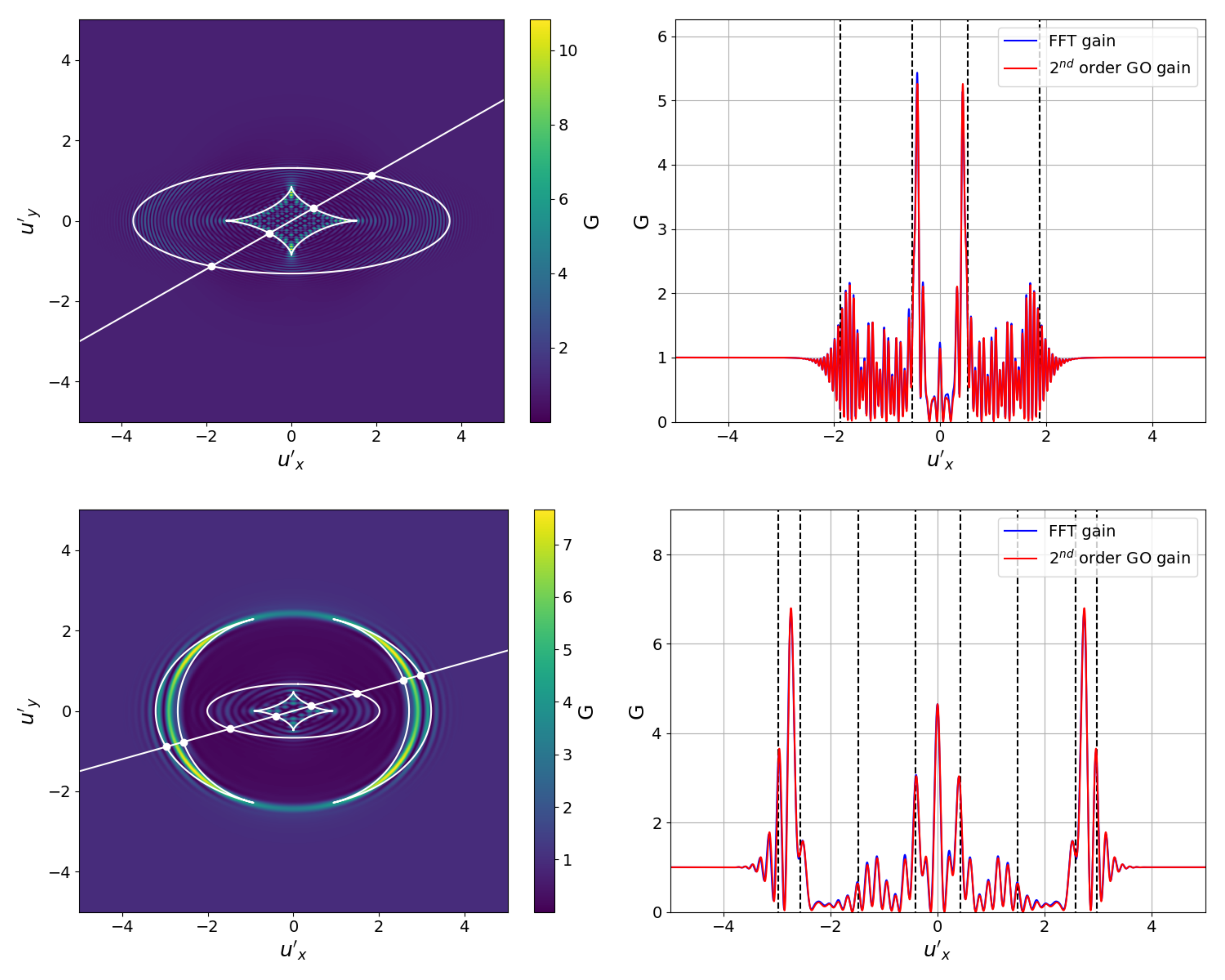}
\par\end{centering}
\caption{\label{fig: 4}Comparison of the gains obtained from a full numerical
solution of the KDI and second order geometric optics. The left column
shows color maps of the gain obtained by solving the KDI via the FFT.
The white circles correspond to caustic curves, and the straight white
line shows the path of the observer through the $u'$ plane. The right
column shows the gain along this path as calculated via the FFT method
(blue) and via second order geometric optics (red). The points of
intersection between the caustics and the observer path are marked
by white points in the left column and by dashed vertical black lines
on the plots in the right column. The top panels shows results for
an underdense elliptical Gaussian lens with $\psi(\boldsymbol{u})=\exp\left(-u_{x}^{2}-u_{y}^{2}\right)$,
lens phase $\phi_{0}=100$ rad, and lens scales $a_{x}=2\times10^{-2}$
AU and $a_{y}=3\times10^{-2}$ AU. The bottom panel corresponds to
an overdense ring-like lens with $\psi(\boldsymbol{u})=2.72\left(u_{x}^{2}+u_{y}^{2}\right)\exp\left(-u_{x}^{2}-u_{y}^{2}\right)$,
lens phase $\phi_{0}=-30$ rad, and lens scales $a_{x}=2\times10^{-2}$
AU and $a_{y}=3\times10^{-2}$ AU. The frequency of observation is
$\nu=0.8$ GHz, $d_{so}=1$ kpc, $d_{sl}=0.5$ kpc for both the top
and bottom panels. The central caustic at the center of both $u'$
planes in the left column occur because $a_{x}\protect\neq a_{y}$,
and is known as a structurally stable caustic of primary aberration
(\citealt{Berry&Upstill1980}).}
\end{figure*}

Matching this to Eq. \ref{eq: 33}, we obtain two sets of equations
that can be used to determine $g_{1}$, $g_{2}$, $\chi$, and $\xi$
in terms of the geometrical optics amplitudes $A_{j}$, and the phases
$\beta_{j}^{r}$. The first set is

\begin{eqnarray}
A_{1} & = & \frac{1}{\sqrt{2}}\left[g_{1}(-\xi)^{\nicefrac{-1}{4}}+ig_{2}(-\xi)^{\nicefrac{1}{4}}\right]\nonumber \\
A_{2} & = & \frac{1}{\sqrt{2}}\left[g_{1}(-\xi)^{\nicefrac{-1}{4}}-ig_{2}(-\xi)^{\nicefrac{1}{4}}\right]\label{eq: 37} .
\end{eqnarray}

Solving for $g_{1}$ and $g_{2}$ gives
\begin{eqnarray}
g_{1} & = & \frac{1}{\sqrt{2}}(A_{1}+A_{2})(-\xi)^{\nicefrac{1}{4}}\nonumber \\
g_{2} & = & \frac{i}{\sqrt{2}}(A_{1}-A_{2})(-\xi)^{\nicefrac{-1}{4}}\label{eq: 38} .
\end{eqnarray}

The second set of equations is
\begin{eqnarray}
\chi+\gamma & = & \beta_{1}^{r}\nonumber \\
\chi-\gamma & = & \beta_{2}^{r}\label{eq: 39},
\end{eqnarray}
which leads to
\begin{eqnarray}
\chi & = & \frac{1}{2}(\beta_{1}^{r}+\beta_{2}^{r})\nonumber \\
\xi & = & -\left[\frac{3}{4}\left(\beta_{1}^{r}-\beta_{2}^{r}+\frac{\pi}{2}\right)\right]^{\nicefrac{2}{3}}\label{eq: 40}.
\end{eqnarray}

Putting everything together, we obtain the uniform asymptotic for
the fold caustic's bright side,
\begin{multline}
\varepsilon_{{\rm bright}}(\boldsymbol{u'},\nu)=\sqrt{\pi}e^{i\chi}\left[(A_{1}+A_{2})(-\xi)^{\nicefrac{1}{4}}\Ai(\xi)\right.\\
\left.+i(A_{1}-A_{2})(-\xi)^{\nicefrac{-1}{4}}\Ai'(\xi)\right]\label{eq: 41} .
\end{multline}

The ambiguity in the labeling is resolved by the condition $\beta_{1}-\beta_{2}+\pi/2>0$.
The merging rays will in general have opposite parities, with $\beta_{j}^{r}-\Phi_{j}=0$
for one ray and $\beta_{j}^{r}-\Phi_{j}=\pm\pi/2$ for the other,
so this condition is equivalent to $\Phi_{1}-\Phi_{2}>0$. Note that
even though $A_{1}$ and $A_{2}$ diverge as they approach the singularity,
the quantity $(A_{1}+A_{2})(-\xi)^{\nicefrac{1}{4}}$ goes to a finite
limit, because $\xi\rightarrow0$ at the caustic. By the same token,
although $(-\xi)^{\nicefrac{-1}{4}}$ goes to infinity at the singularity,
the quantity $(A_{1}-A_{2})(-\xi)^{\nicefrac{-1}{4}}$ does not, because
$A_{1}-A_{2}\rightarrow0$.

At the caustic's dark side, we know from \S\ref{subsec: 3.1} that
far from the singularity, the geometrical optics field reduces to
that of a single complex ray, $\varepsilon_{+}^{c}=Ae^{-\Phi_{i}}e^{i\beta_{+}^{c}}$.
Therefore, our ansatz no longer contains $\Ai'(\xi)$.
Instead, we have that
\begin{equation}
\varepsilon_{{\rm dark}}(\boldsymbol{u'},\nu)=\sqrt{2\pi}e^{i\chi}g_{0}\Ai(\xi)\label{eq: 42} .
\end{equation}

The asymptotic of $\Ai(\xi)$ for large positive argument is
\begin{equation}
\Ai(\xi)\approx\frac{\exp\left(-\frac{2}{3}\xi^{\nicefrac{3}{2}}\right)}{2\xi^{\nicefrac{1}{4}}\sqrt{\pi}}\label{eq: 43}.
\end{equation}

Matching coefficients as before, we obtain $\chi=\beta_{+}^{c}$,
$\xi=\left[\frac{3}{2}\Phi_{i}\right]^{\nicefrac{2}{3}}$, and $g_{0}=A\xi^{\nicefrac{1}{4}}\sqrt{2}$.
Thus, 
\begin{equation}
\varepsilon_{{\rm dark}}(\boldsymbol{u'},\nu)=2\sqrt{\pi}e^{i\beta_{+}^{c}}A\xi^{-\nicefrac{1}{4}}\Ai(\xi)\label{eq: 44}.
\end{equation}
 Again, even though at the caustic $A\rightarrow\infty$, the expression
does not diverge because $\xi\rightarrow0$ at the same point.

\subsection{Uniform asymptotics in plasma lensing}

For the present case of plasma lenses and astrophysical distances,
the idealized situation presented above involving two real images
on the caustic's bright side and no real images on its dark side does
not actually occur, as the lens is not opaque and the immense distances
allow the initial cone of emitted radiation to grow to a size much
larger than that of the lens by the time the two encounter each other.
Thus, Eqs. \ref{eq: 41} and \ref{eq: 44} cannot be applied directly
as given: there will always be at least one real ray involved in the
description of the field, and the total number of rays will always
be odd. The more general way of dealing with such a situation would
be to implement the uniform asymptotic for the next catastrophe in
the series, the cusp. The canonical integral in that case is

\begin{equation}
I_{{\rm cusp}}(\xi_{1},\xi_{2})=\frac{1}{\sqrt{2\pi}}\int_{-\infty}^{\infty}dt\exp\left[i\left(\xi_{1}t+\frac{\xi_{2}t^{2}}{2}-\frac{t^{4}}{4}\right)\right]\label{eq: 45},
\end{equation}
which is related to the Pearcey integral $P(\xi_{1},\xi_{2})$ (\citealt{Pearcey1946})
by the relationship $I_{{\rm cusp}}(\xi_{1},\xi_{2})=P^{*}(-\sqrt{2}\xi_{1},-\xi_{2})/\sqrt{\pi}$.
The observer sees three images in the bright side and one image in
the dark side, which is exactly what happens for the Gaussian lens
analyzed in Figure \ref{fig:2}. The corresponding ansatz for the
bright side of the caustic would then be
\begin{equation}
\varepsilon(\boldsymbol{u'},\nu)=e^{i\chi}\left[g_{1}I_{{\rm cusp}}(\xi_{1},\xi_{2})+g_{2}\frac{\partial I_{{\rm cusp}}}{\partial\xi_{1}}+g_{3}\frac{\partial I_{{\rm cusp}}}{\partial\xi_{2}}\right]\label{eq: 46} .
\end{equation}

Finding the unknown parameters $g_{j}$, $\xi_{j}$, and $\chi$,
however, is not possible via implementation of the same matching procedure
we used above for the fold caustic. One reason for this is that the
asymptotic forms of the Pearcey integral are much more complicated
than those of the Airy integral (\citealt{Paris1991}). Instead, the
correct strategy involves obtaining systems of equations for the relevant
quantities by exploiting the correspondence between the phase function
of the canonical integral and the phase function of the KDI at the
stationary points, as described in detail by \citet{Connor1973b}
and \citet{Katsaounisetal2001}. Unfortunately, in the case of cusps
and higher order catastrophes, it is not possible to express all the
unknown parameters as a function of the geometrical optics quantities
in a simple form.

For practical purposes, however, this is rarely necessary. Cusps correspond
to points in which three solutions of the lens equation merge. These
points are connected to each other by curves which correspond to fold
catastrophes, where only two images merge. Far from these cusp points,
Eq. \ref{eq: 46} can be written as the sum of the uniform asymptotic
for the fold caustic and the regular geometric optics contribution
from each of the $n$ images not involved in the formation of the
fold lines. This also holds for higher order catastrophes. Thus, as
long as we are not too close to catastrophes of higher order, the
total field can be written as

\begin{equation}
\varepsilon(\boldsymbol{u'},\nu)=\varepsilon_{{\rm fold}}+\sum_{j=1}^{n}A_{j}e^{i\beta_{j}^{r}}\label{eq: 47},
\end{equation}
where $\varepsilon_{{\rm fold}}$ is given by Eq. \ref{eq: 41} or
Eq. \ref{eq: 44} depending on whether we are at the caustic's dark
side or bright side.

\begin{figure*}[t]
\begin{centering}
\includegraphics[width=0.9\textwidth]{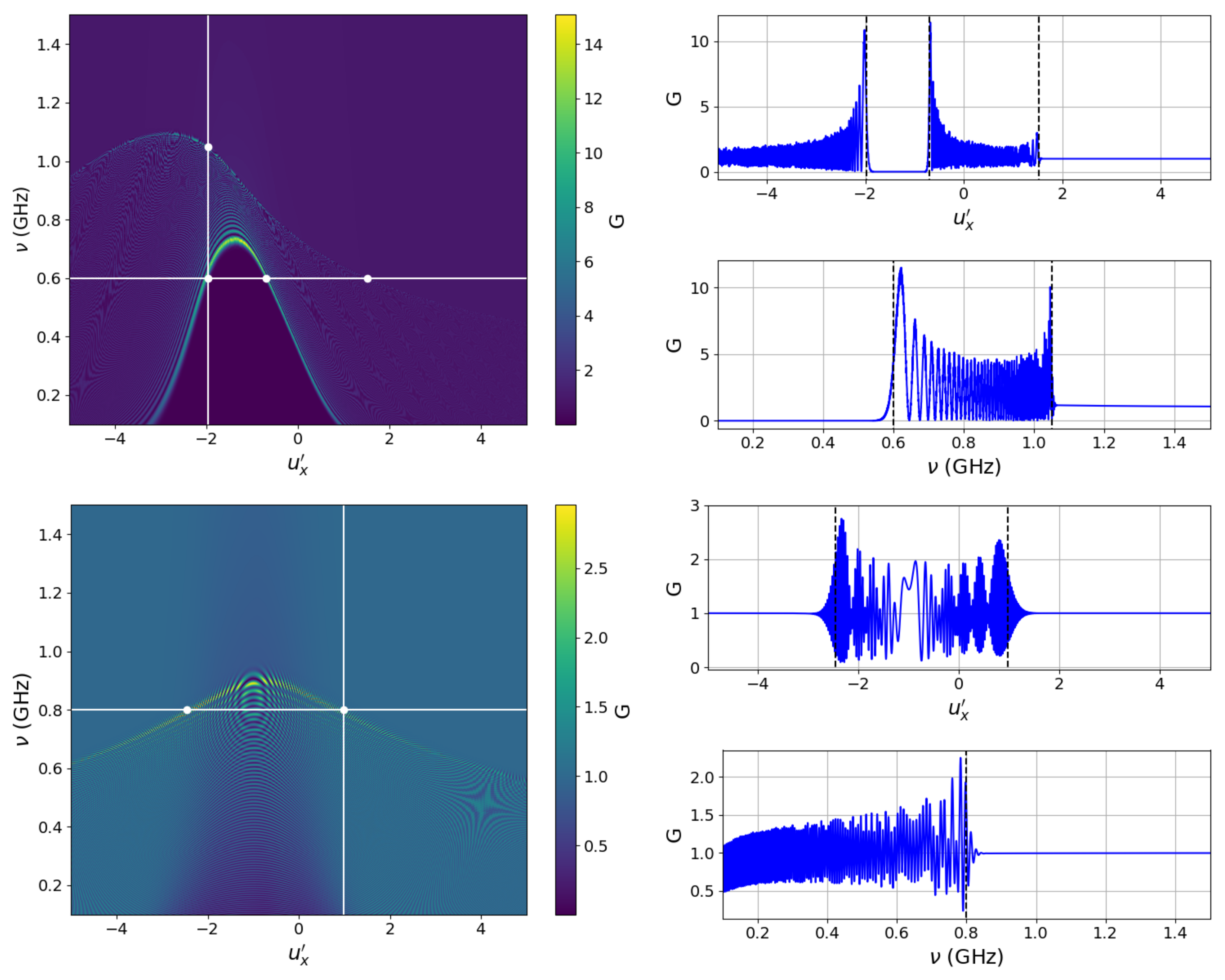}
\par\end{centering}
\caption{\label{fig: 5}Sections of dynamic spectra and slices across them
for overdense and underdense perturbed Gaussian lenses with $\psi(\boldsymbol{u})=\exp(-u_{x}^{2}-u_{y}^{2})\left\{ 1-A\left[\sin(Bu_{x})+\sin(Bu_{y})\right]\right\} $
and different ${\rm DM}_{\ell}$ magnitudes. The left column shows
the two dimensional spectrum for both lenses, with the top row corresponding
to the overdense lens and the bottom row to the underdense lens. The
vertical and horizontal lines correspond to the slices across the
spectra plotted in the right column. Caustic intersections are marked
by white dots in the left column plots and by dashed black lines in
the right column plots. The overdense lens has a maximum column density
of ${\rm DM}_{\ell}=10^{-4}$ pc cm\protect\textsuperscript{-3} and
lens scales of $a_{x}=0.1$ AU and $a_{y}=0.2$ AU, whereas the underdense
lens has ${\rm DM}_{\ell}=-10^{-5}$ pc cm\protect\textsuperscript{-3},
and $a_{x}=a_{y}=0.04$ AU. Both lenses have perturbation parameters
$A=1.5\times10^{-2}$ and $B=5$, source-observer distance $d_{so}=1$
kpc, and source-lens distance $d_{sl}=0.5$ kpc. The path through
the $u'$ plane in both cases is a straight line with slope $m=0.5$
and y-intercept $n=2.5$.}
\end{figure*}

Figure \ref{fig: 4} shows a comparison between the gain obtained
from the FFT and that obtained using the uniform asymptotic formulas
for a slice across the $u'$ plane, for two different lens shapes
$\psi$ and lens phases $\phi_{0}$. Unlike the case of the circular
Gaussian lens with positive ${\rm DM}_{\ell}$ depicted in Figure
\ref{fig:2}, the lenses in these figures show cusps as well as folds.
In Figure \ref{fig: 4}, both the elliptical Gaussian with ${\rm DM}_{\ell}<0$
(top panels) and the ring-like lens with ${\rm DM}_{\ell}>0$ (bottom
panels) show fold lines interrupted by cusp points at which three
roots merge and the curvature of the fold lines is reversed. The number
of images that can be seen varies depending on the position in the
$u'$ plane and the type of lens. For the negative ${\rm DM}_{\ell}$
elliptical Gaussian, the observer sees one image in the dark side
of the outer caustic zone, three images after crossing the outer caustic
boundary, and five images in the central caustic. For the ring-like
lens in the bottom panels, the number of images is equal to one outside
the caustic zones, three inside the mirrored crescent shaped caustics
and in between the two central caustic curves, and five at the center.
Other lens shapes can show larger numbers of images and catastrophes
of higher order. Some examples are shown in Appendix \ref{sec: Appendix D}.

\subsection{Advantages of second order geometric optics}

As long as $|\phi_{o}|\gg1$, second order geometric optics is able
to produce remarkably accurate results. Unlike the FFT method, it
can be implemented for essentially arbitrary values of $a_{x}$, $a_{y}$,
and $\phi_{0}$ without difficulty. We have applied the second order
approach only to the case of slices across the $u'$ plane at a fixed
frequency of observation, but the equations for the field hold identically
if we were to vary any of the parameters present in the phase function
Eq. \ref{eq: 18}. Thus, we can use second order geometric optics
to produce accurate plots of the gain as a function of $\nu$ at a
fixed position in the $u'$ plane. Even for small values of the lens
scales, constructing such a plot using the FFT would be extremely
computationally expensive, as it would require performing two dimensional
FFTs at each frequency of observation.

Using the concepts developed so far, we can construct sections of
the dynamic spectrum of a lens event, at least for the case in which
these show no cusps. Plots of the gain as a function of position along
a line in the $u'$ plane at a single frequency will then correspond
to horizontal slices of the dynamic spectrum, whereas plots of the
gain as a function of $\nu$ at fixed $u'$ coordinates will correspond
to vertical slices. This is illustrated further in Figure \ref{fig: 5}.
From the figure, it is also apparent that larger magnitudes of the
maximum column density $\left|{\rm DM}_{\ell}\right|$ induce faster
oscillations in the gain, and the contributions from complex rays
in the shadow sides of caustics become less important.

\section{Astrophysical applications}\label{sec: 4}

\subsection{TOA perturbations}\label{subsec: 4.1}

One of the important potential effects of plasma lensing, in particular
with regards to its consequences to pulsar timing, is the issue of
perturbations in pulse arrival times. The importance of these potential
perturbations has been clear for a long time (see, e.g. \citealt{Cordes&Wolszcan1986,Cordesetal1986})
and has resurfaced more recently given the potential of PTAs to detect
low frequency gravitational waves (\citealt{Cordes&Shannon2010,Cordesetal2016})
and in the context of FRBs (\citealt{Cordesetal2017,Dai&Lu2017}).
Our analysis will rely on examples that use parameters that are more
likely to be relevant for pulsar timing, where the resulting perturbations
are in the order of microseconds, and the distances place the source
and lens inside the Milky Way galaxy. Nevertheless, the same concepts
can be applied to the FRB case by increasing the distances, the lens
sizes, and the magnitude of the maximum dispersion measure perturbations.

\subsubsection{Geometry and dispersion}

Refraction due to plasma lensing invariably introduces a geometric
delay into the time of arrival of radiation, independently of whether
the lensing effect is produced by an underdensity or an overdensity
in the interstellar medium. By Fermat's principle, an unlensed ray
will travel in a straight line from the source to the observer, and
lensing introduces a deviation from this straight path. Referring
to the geometry of Figure \ref{fig:1}, we can write the magnitude
of the geometric delay $\Delta t_{geo}$ as
\begin{equation}
\Delta t_{{\rm geo}}=t_{g_{x}}(u_{x}-u_{x}')^{2}+t_{g_{y}}(u_{y}-u_{y}')^{2}\label{eq: 48},
\end{equation}
where the $t_{g_{x,y}}=a_{x,y}^{2}d_{so}/2cd_{sl}d_{lo}$ are the geometrical
delay coefficients along the $u_{x,y}$ axes. The location of images
in the $u$ plane is determined by the coordinates in the $u'$ plane
and the lens equation, so for a given image located at $\boldsymbol{u}=\boldsymbol{u_{j}^{0}}$,
we can express the geometric delay as
\begin{equation}
\Delta t_{{\rm geo}}=t_{g_{x}}\alpha_{x}^{2}\psi_{10}^{2}+t_{g_{y}}\alpha_{y}^{2}\psi_{01}^{2}\label{eq: 49}.
\end{equation}

\begin{figure*}[t]
\begin{centering}
\includegraphics[width=0.95\textwidth]{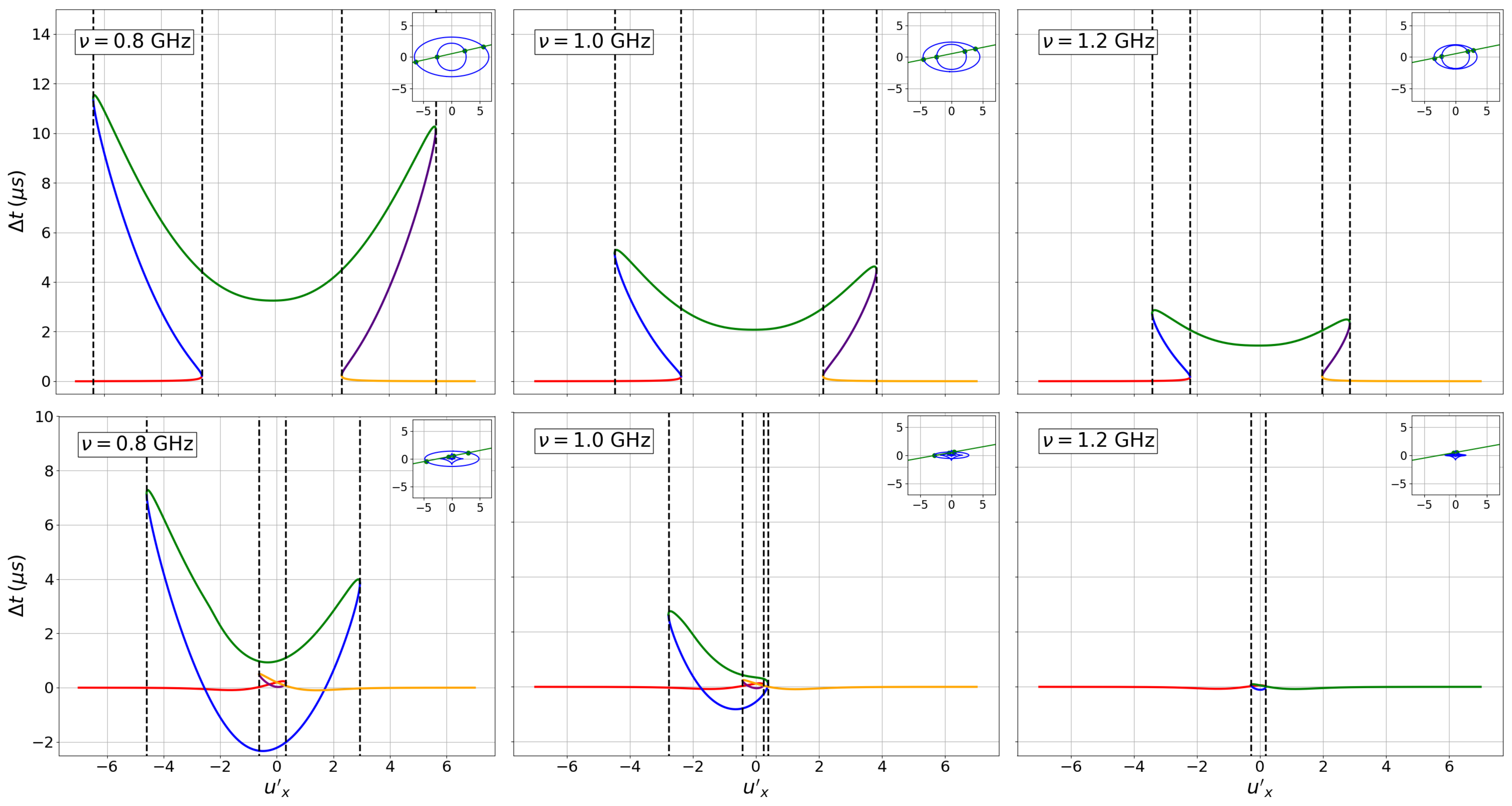}
\par\end{centering}
\caption{\label{fig: 6}
Timing perturbations of for pulses corresponding to different images as a function of observer position for overdense (top row) and underdense
(bottom row) lenses with ${\rm DM}_{\ell}=\pm5\times10^{-4}$ pc
cm\protect\textsuperscript{-3}. Different frames in each row correspond to different frequencies of observation. Both overdense and underdense lenses have a Lorentzian shape
with $\psi(\boldsymbol{u})=\nicefrac{1}{\left[\left(u_{x}^{2}+u_{y}^{2}\right)^{2}+1\right]}$
and lens scales $a_{x}=0.25$ AU, $a_{y}=0.4$ AU. The distances used
were $d_{so}=1$ kpc and $d_{sl}=0.5$ kpc, and the path through the
$u'$ plane has slope $m=0.2$ and y-intercept $n=0.5$. The subplot
in the top corner of each subpanel shows the (blue) caustic curves
in the $u'$ plane for the corresponding frequency of observation,
together with the (green) path of the observer through the $u'$ plane.
The different colors in the $\Delta t$ vs $u_{x}'$ plots trace the timing perturbation for each individual image.}
\end{figure*}

Independently of the geometric delay, the lens will also introduce
a dispersive perturbation in pulse arrival time due to the plasma's
effect on the radiation's group velocity $v_{g}$. For a cold plasma,
$v_{g}v_{p}=c^2$, which means that the dispersive perturbation in the
TOA is given by
\begin{eqnarray}
\Delta t_{{\rm DM}} & = & \frac{cr_{e}{\rm DM}_{\ell}}{2\pi\nu^{2}}\psi(\boldsymbol{u})\nonumber \\
 & = & 4.149\,\text{ms}\times\frac{{\rm DM}_{\ell}}{\nu^{2}}\psi(\boldsymbol{u})\label{eq: 50},
\end{eqnarray}
where the second equality applies for a ${\rm DM}_{\ell}$ in units
of pc cm\textsuperscript{-3} and $\nu$ in GHz. If ${\rm DM}_{\ell}>0$,
the dispersive perturbation will introduce a TOA delay, as the column
density of electrons along the line of sight will increase\footnote{Of course, it is possible to have a lens function $\psi$ that is
both positive and negative depending on $\boldsymbol{u}$, such as
$\psi(\boldsymbol{u})=\sin(u_{x})+\sin(u_{y})$. Thus, this statement
is correct only for lens realizations that have $\psi>0$ for all
$\boldsymbol{u}$, which is the case for all the examples shown throughout
this work.}. On the other hand, if ${\rm DM}_{\ell}<0$, the lens will constitute
a ``pinhole'' in the interstellar medium, and radiation passing
through the lens will experience less of a dispersive delay than radiation
traveling outside of it.

The total TOA perturbation for each image $\Delta t_{j}$ is simply
the sum of the geometric and dispersive perturbations, 
\begin{equation}
\Delta t_{j}=\Delta t_{{\rm geo}}^{j}+\Delta t_{{\rm DM}}^{j}\label{eq: 51}.
\end{equation}

When ${\rm DM}_{\ell}>0$, both perturbations are positive, and the
total TOA perturbation will be positive for any combination of parameters,
frequency of observation, and position in the $u'$ plane. On the
other hand, when ${\rm DM}_{\ell}<0$, $\Delta t_{j}$ can be either
positive or negative depending on the relative magnitudes of $\Delta t_{{\rm geo}}^{j}$
and $\Delta t_{{\rm DM}}^{j}$. For an observer close to the origin
of the $u'$ plane and a lens with a maximum dispersion measure perturbation
at the center of the lens plane, the maximum TOA advance will occur
for solutions to the lens equation that are within the $u$ plane's
central region, since at these points the geometric delay will be
minimum and the dispersive advance will be maximum. For a fixed position
in the $u'$ plane, the magnitude of the geometric perturbation for
an individual image will decrease as $\nu^{-4}$ (\citealt{Rickett1990}),
whereas the dispersive delay will decrease as $\nu^{-2}$, which means
that dispersive delays will dominate geometric perturbations at large
frequencies. Geometric delays will grow as a function of the lens
size, but larger lenses do not necessarily increase the maximum dispersion
measure perturbation, so geometric delays acquire more significance
as the lens size grows and ${\rm DM}_{\ell}$ stays constant. In general,
the magnitude of the total TOA perturbation per image decreases as
a function of frequency.

Figure \ref{fig: 6} shows a sequence of plots of $\Delta t$ along
a path through the $u'$ plane for frequencies of observation $0.8$,
$1.0$, and $1.2$ GHz for overdense and underdense Lorentzian lenses
with ${\rm DM}_{\ell}=\pm5\times10^{-4}$ pc cm\textsuperscript{-3},
which gives a lens phase $\phi_{o}$ for each of the frequencies of
$\sim\mp1.6\times10^{4}$ rad, $\mp1.3\times10^{4}$ rad, and $\mp1.1\times10^{4}$
rad, respectively. For the overdense lens sequence in the top panels,
both the geometric and dispersive perturbations are positive. At $\nu=0.8$
GHz, the geometric contribution dominates over the dispersive contribution,
as is apparent by the facts that one, the maximum TOA delay occurs
far from the origin of the $u'$ coordinate system, where the geometric
perturbation is larger than the dispersive perturbation, and two,
the minimum delay in the caustic zone occurs at the origin, where
the dispersive delay is maximum and the geometric delay is minimum.
Outside of the caustic region, the delay is negligible. As we increase
the frequency, it can be seen that the difference between the minimum
delay at the center and the maximum delay at the edges of the caustic
zone becomes less noticeable, as the magnitude of the geometric delay
decreases faster than that of the dispersive delay. The maximum number
of images produced in the case of the overdense lens is three, and
the caustic pattern is very similar to that of an overdense two dimensional
Gaussian like the one depicted in Figure \ref{fig:2}.

The bottom panels, corresponding to the lens with ${\rm DM}_{\ell}<0$,
show a different sequence. This time the maximum number of images
(five) is seen along the section of the observer's path through $u'$
that is closer to the center of the caustic region, and the caustic
curves form cusps as well as folds. The dispersive perturbation is
now negative, and is able to overpower the geometric delay only in
regions close to the origin, where the geometric delay is at a minimum.
Nevertheless, only one of the five images actually shows a TOA advance.

In both the overdense and the underdense case, we see that the total
magnitudes of the perturbations decrease as a function of frequency,
and almost no lensing effects are apparent at 1.2 GHz, although this
is more dramatic for the underdense lens than for the overdense one.
Both the size of the caustic zone and the distance between each of
the caustic curves decrease as as a function of frequency, because
of the weakening of the lens's refractive power.

\subsubsection{Telescope observations of TOA perturbations during a lensing event}

The examples from the previous section apply only to the unrealistic
case of measurements performed at an infinitely narrow frequency band,
and ignore the fact that in general a telescope will be unable to
resolve individual images. In reality, the incident electric field
$E(t)$ is sampled as a function of time by the telescope's receiver,
and individual pulse shapes are constructed by taking the Fourier
transform of $E(t)$, $\tilde{E}(\nu)=\int dt\,E(t)e^{-2\pi i\nu t}$
and transforming back after filtering $\tilde{E}(\nu)$ with a bandpass
of bandwidth $\Delta\nu_{r}$ centered on frequency $\nu_{0}$. Then,
the electric field measured by the telescope across a single band
$E_{{\rm band}}$ can be written as (\citealt{Cordes&Wasserman2016})
\begin{equation}
E_{{\rm band}}(t,\nu_{0};\Delta\nu_{r})=\int_{\nu_{0}-\Delta\nu_{r}/2}^{\nu_{0}+\Delta\nu_{r}/2}d\nu\,\tilde{E}(\nu)e^{2\pi i\nu t}\label{eq: 52} .
\end{equation}

The pulse profile for the band can then be constructed by taking the
square modulus of Eq. \ref{eq: 52}. The effects of lensing can be
quantified as follows. Let an unlensed pulse be described by a normalized
electric field $V_{0}(t)$ and Fourier transform $\tilde{V}_{0}(\nu)$.
Then, the lensed pulse over a band $V_{{\rm band}}$ will be given
by
\begin{align}
V_{{\rm band}}(\boldsymbol{u'},t,\nu_{0};\Delta\nu_{r}) & =\int_{\nu_{0}-\Delta\nu_{r}/2}^{\nu_{0}+\Delta\nu_{r}/2}d\nu\,\tilde{V}_{0}(\nu)\varepsilon(\boldsymbol{u'},\nu)e^{2\pi i\nu t}\label{eq: 53},
\end{align}
where $\varepsilon(\boldsymbol{u'},\nu)$ is the normalized scalar
field from the monochromatic KDI, Eq. \ref{eq: 17}. As we showed
in \S\ref{sec: 3}, we can accurately and efficiently solve the KDI
using second order geometric optics, by expressing $\varepsilon(\boldsymbol{u'},\nu)$
as a sum of terms of the same form as Eq. \ref{eq: 47}. In practice,
$V_{0}(t)$ looks like modulated white noise, and the processing of
the data captured by the telescope will involve heterodyning to baseband,
coherently dedispersing, and the folding of multiple pulses to obtain
a better signal to noise ratio. Nevertheless, in the context of numerical
simulations, we can get an idea about how lensing events will show
up in our data by regarding the unlensed pulse as a unit impulse at
$t=0$, $V_{0}(t)=\delta(t)$. Then, $\tilde{V}_{0}(\nu)=1$, and
the deviation of $V_{{\rm band}}$ from $V_{0}$ will be exclusively
due to the characteristics of the lens and the observing position
$\boldsymbol{u'}$. We can mimic how the perturbation will look in
real data by convolving $I=\left|V_{{\rm band}}\right|^{2}$ with
a suitable pulse template and adding white noise. The TOA perturbation
for each band can be calculated afterwards using PyPulse\footnote{Lam, M. T., 2017, PyPulse, Astrophysics Source Code Library, record
ascl:1706.011}.

\begin{figure*}[t]
\begin{centering}
\includegraphics[width=0.95\textwidth]{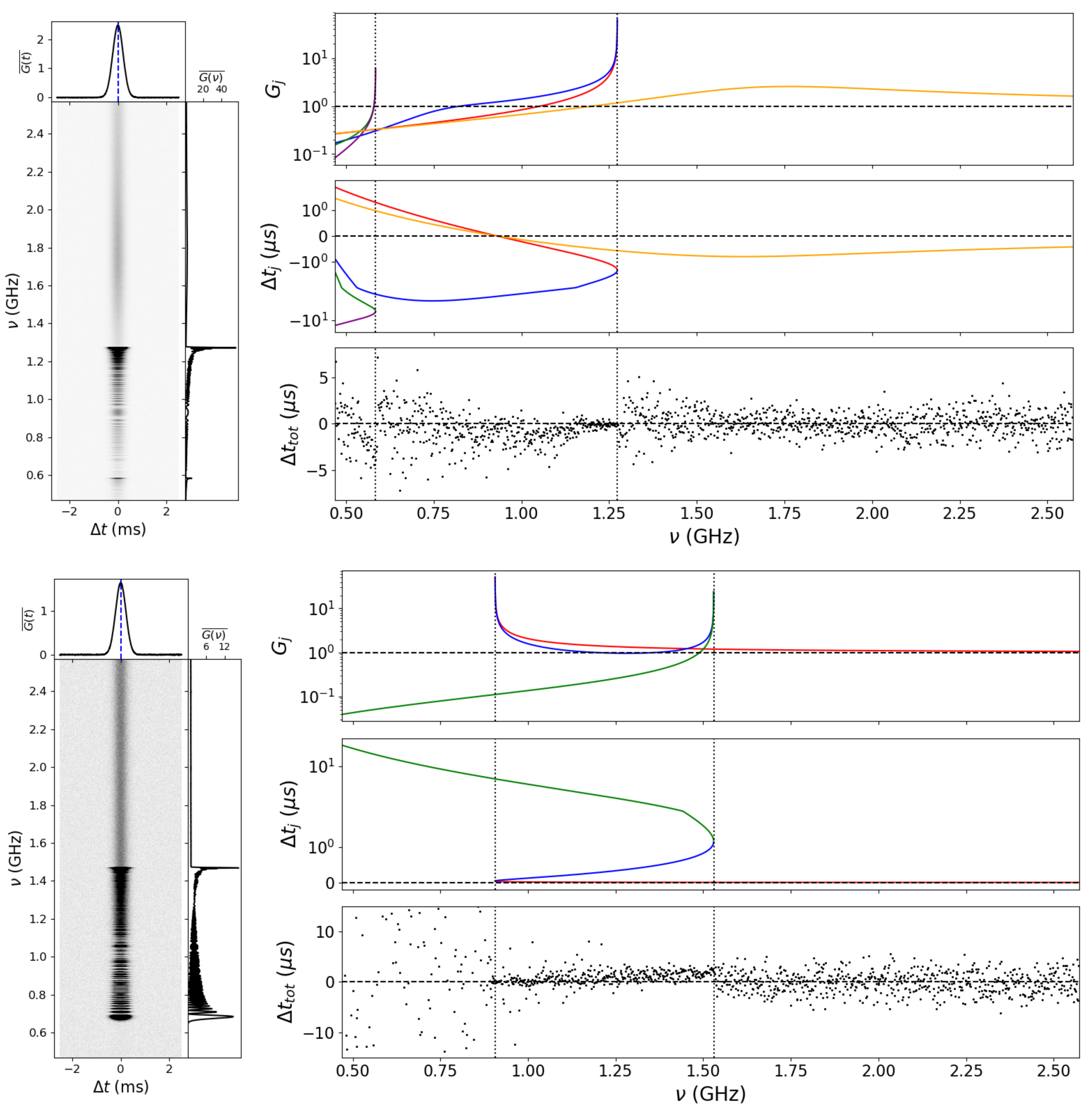}
\par\end{centering}
\caption{\label{fig: 7}Pulse dynamic spectra, individual image gains and TOA
perturbations, and combined TOA perturbation as a function of frequency
for a single epoch of observation for an underdense lens (top panel)
and an overdense lens (bottom panel). The lens in the top panel has
Gaussian shape $\psi(\boldsymbol{u})=\exp(-u_{x}^{2}-u_{y}^{2})$,
${\rm DM}_{\ell}=-7\times10^{-4}$ pc cm\protect\textsuperscript{-3},
$d_{so}=1$ kpc, $d_{sl}=0.5$ kpc, $a_{x}=0.5$ AU and $a_{y}=1.1$
AU, and the epoch corresponds to a position in the $u'$ plane with
coordinates $\boldsymbol{u'}=(0.1,0.1)$. The lens in the bottom panel
has shape $\psi(\boldsymbol{u})=\exp\left[-\left(u_{x}^{2}+u_{y}^{2}\right)^{2}\right]$,
${\rm DM}_{\ell}=10^{-3}$ pc cm\protect\textsuperscript{-3}, $d_{so}=2$
kpc, $d_{sl}=1.5$ kpc, $a_{x}=0.8$ AU, $a_{y}=1.1$ AU, and $\boldsymbol{u'}=(-1.5,-0.55)$.
The pulse profile contains 2048 bins, and the pulse repetition period
is $T=5$ ms, giving an integration time $\Delta t_{{\rm int}}\approx2.44$
$\mu\text{s}$. The channel bandwidth is $\Delta\nu_{r}=1.5$ MHz.
We use a Gaussian template to model the pulse shape.}
\end{figure*}

Figure \ref{fig: 7} shows numerically simulated pulse dynamic spectra,
individual image gains and TOA perturbations, and combined TOA perturbations
as a function of frequency for a single epoch of observation (fixed
$\boldsymbol{u'}$). In the top panel, lensing occurs as a result
of an underdensity with Gaussian shape, $\psi(\boldsymbol{u})=\exp(-u_{x}^{2}-u_{y}^{2})$,
whereas in the bottom panel, the lens is overdense with shape $\psi(\boldsymbol{u})=\exp\left[-\left(u_{x}^{2}+u_{y}^{2}\right)^{2}\right]$.
The parameters used, listed in the figure's caption, lead to caustic
formation in both cases. The effects of the lenses in pulse TOAs vary
dramatically as a function of frequency, especially close to the caustics,
where the signal to noise ratio can be observed to increase as the
image magnification becomes large, and sharp discontinuities arise
as images appear or disappear.

The case of the underdense lens is especially complex due to the fact
that different images can have either positive or negative TOA perturbations,
meaning that the overall pulse TOA can be delayed or advanced depending
on the frequency and epochs of observation, the lens parameters, and
the lens shape. It is a generic feature of underdense lenses that
the number of images tends to decrease as a function of frequency,
and thus we expect to be able to observe multiple imaging events more
often at low frequencies than at large frequencies. The size of the
multiple images regions, as well as the distance between caustics,
will in general change as we vary the coordinates in the $u'$ plane,
the value of ${\rm DM}_{\ell}$, and the size of the lens.

For the overdense lens in the figure, the region of multiple imaging
follows a region of very large demagnification of a single image,
a behavior that can also be observed in the case of the stochastic
Gaussian lens shown in Figure \ref{fig: 5}, and appears to be generic for
the case of Gaussian-like overdense lenses, although more complicated
lens shapes can lead to other types of behavior. The signal to noise
ratio in the first region is therefore extremely low, and the perturbations
are dominated by white noise. The region of multiple imaging shows
a gradual increase in the TOA delay as a function of frequency, with
the signal to noise ratio increasing as we move closer to the caustic,
after which the lensing effects are minimal. Again, we can see sharp
discontinuities in the behavior of the perturbations at both caustic
points.

\subsection{Dispersion measure perturbations}\label{subsec: 4.2}

Modern pulsar timing models and pulsar timing packages like TEMPO
and TEMPO2 operate on the assumption that the frequency dependent
delay of incoming radiation is purely dispersive, with the total delay
being given by

\begin{equation}
\Delta t=4.149\,\text{ms}\times\frac{{\rm DM}}{\nu^{2}}\label{eq: 54},
\end{equation}
where DM is in standard units of pc cm\textsuperscript{-3} and $\nu$
is in GHz. Physically, DM corresponds to the total integrated column
density of electrons along the line of sight between the Earth and
the pulsar. As discussed in the previous section, a lens changes the
dispersive contribution depending on its characteristic shape and
the parameter ${\rm DM_{\ell}}$, but also introduces a geometric
perturbation in the TOAs due to refraction. These perturbations will
be different for each image of the source for the cases in which the
lensing is strong enough for ray crossings to occur. Thus, during
a strong lensing event like the ones we have analyzed in this work,
the expected $\nu^{-2}$ relationship for the group delay will not
in general hold. Furthermore, we would expect that attempts at finding
the best value of DM according to Eq. \ref{eq: 54} will yield different
best fit values and different deviations from the expected $\nu^{-2}$
scaling depending on the frequency band. This follows from the fact
that the nature of the frequency dependence of the perturbations due
to the lens can change drastically as a caustic is crossed, as illustrated
in Figure \ref{fig: 7}. This also means that a lensing event will
not necessarily show up in the data as an increase in the $\nu^{-4}$
dependence of the residuals, except in cases in which the frequency
band across which the data is being analyzed contains only a single
image. A more sophisticated analysis, taking into account the details
involved in the operational determination of DM and the way it changes
in time, as described in \citet{Keithetal2013}, is outside the scope
of this work.

\section{Summary and conclusions}\label{sec: 5}

We have built on previous works that have studied the phenomenon of
astrophysical plasma lensing in the context of ESEs, scintillations,
and FRBs by developing a more general formalism that applies to two
dimensional plasma lenses formed by both underdensities and overdensities
in the ISM, and that can be used to study and predict the many possible
ways in which lensing can affect observational quantities such as
pulse intensities and TOAs. We showed that the geometrical optics
method commonly employed in previous works to construct lensed light
curves is unable to properly describe the fluctuations in the gain
due to the interference between multiple source images, and is also
unable to properly describe the gain within caustic zones.

By incorporating elements of catastrophe theory and the study of uniform
asymptotic approximations of highly oscillatory integrals, we have
developed an enhanced version of geometric optics that is able to
account for such oscillatory features, and that does not break down
at caustic curves in which two geometric optic images merge. We showed
how this type of geometric optics can be successfully leveraged to
construct the flux perturbations due to a variety of lens shapes and
sizes, overcoming some of the limitations of other numerical approaches.
We also apply some elements of this approach to characterize the possible
form of TOA perturbations due to lensing events.

Our results indicate that there are many ways in which lensing effects
can present themselves to an observer, depending on the lens shape,
the magnitude of the electron density's departure from the surrounding
ISM, whether this departure acquires the form of an overdensity or
underdensity, and a series of other parameters such as the lens size,
distances, and the frequencies of observation. The two dimensional
model also adds an important degree of freedom in the form of the
observer's path through the $u'$ plane, something that cannot be
correctly accounted for by one dimensional models. This extra degree
of freedom also leads to the appearance of higher order diffraction
catastrophes in parameter space that our approach is presently unable
to accurately model. We expect to solve this problem in future work,
as the successful implementation of uniform asymptotic methods for
catastrophes like the cusp can greatly expand the the volume of parameter
space that can be explored accurately in simulations.

Consistent with the results of previous works \citep{Goodmanetal1987,Melrose&Watson2006,Watson&Melrose2006,Stinebringetal2007},
we find that lensing effects tend to be stronger at lower frequencies
since the refractive power of plasma is more pronounced at large wavelengths.
We also find our results for the overdense Gaussian lens to be consistent
with results presented in previous works \citep{Cleggetal1998,Stinebringetal2007,Cordesetal2017,Er&Rogers2017}.
Unlike these studies, however, we also analyze underdense Gaussian
lenses, and find that their observational consequences are dramatically
different from the overdense case. We also apply the uniform asymptotics
approach to other types of lens shapes that have not been explored
in the past.

The increasing accuracy of pulsar timing methods and procedures, as
well as the growing population of pulsars under observation, imply
that relatively rare phenomena like lensing events will be observed
more often, and that their impact on the timing residuals will be
more noticeable. Thus, being able to model such events will become
increasingly more important. We expect to apply the methodology outlined
in this work to establish whether chromatic aberrations such as the
ones reported recently by \citet{Colesetal2015} and \citet{Lametal2018}
are indeed the results of lensing phenomena and, if so, develop a
model of the lensing structures responsible for such occurrences.
The concepts developed here also have direct application to the modelling
of ESEs for sources other than pulsars, and it is possible that lensing
could be part of the explanation for some of the mysteries surrounding
FRBs, which makes future work on this topic all the more important.

\begin{acknowledgements}
G.G. would like to thank Ross Jennings and Michael Lam for helpful correspondence and conversations. The authors acknowledge support from  the NANOGrav Physics Frontiers Center (NSF award 1430284). 
\end{acknowledgements}

\appendix

\section{Solving the KDI using the FFT}\label{sec: Appendix A}

The two dimensional Kirchoff diffraction integral (KDI) introduced
in \S\ref{subsec: 2.3}, gives the normalized wave optics field $\varepsilon$
as a function of the observer coordinates $\boldsymbol{u'}$ by integrating
over an angular spectrum of plane waves,
\begin{equation}
\varepsilon(\boldsymbol{u'},\nu)=\frac{a_{x}a_{y}}{2\pi r_{F}^{2}}\iint d^{2}u\exp(i\Phi),
\end{equation}
where $\Phi$ is the geometric phase,
\begin{equation}
\Phi(\boldsymbol{u'},\boldsymbol{u},\nu)=\frac{1}{2r_{F}^{2}}\left[a_{x}^{2}(u_{x}-u_{x}')^{2}+a_{y}^{2}(u_{y}-u_{y}')^{2}\right]+\phi_{0}\psi(\boldsymbol{u}),
\end{equation}
with $r_{F}$ the Fresnel scale, $a_{x}$ and $a_{y}$ the lens scales,
$\phi_{o}$ the lens strength parameter, and $\psi$ the lens shape.
Given the form of the phase function, the KDI can be written as a
two dimensional convolution integral,
\begin{equation}
\varepsilon(\boldsymbol{u'},\nu)=\iint d^{2}u\,G(\boldsymbol{u}-\boldsymbol{u'},\nu)H(\boldsymbol{u},\nu),
\end{equation}
where
\begin{eqnarray}
G(\boldsymbol{u},\nu) & = & \frac{a_{x}a_{y}}{2\pi r_{F}^{2}}\exp\left[\frac{i}{2r_{F}^{2}}\left(a_{x}^{2}u_{x}^{2}+a_{y}^{2}u_{y}^{2}\right)\right],\\
H(\boldsymbol{u},\nu) & = & \exp\left[i\phi_{o}\psi(\boldsymbol{u})\right].
\end{eqnarray}

From the discrete version of the convolution theorem (\citealt{Schmidt2010}),
we have that
\begin{equation}
\varepsilon(\boldsymbol{u'},\nu)=\mathcal{F}^{-1}\left\{ \mathcal{F}\left[G(\boldsymbol{u},\nu)\right]\cdot\mathcal{F}\left[H(\boldsymbol{u},\nu)\right]\right\}, 
\end{equation}
where $\mathcal{F}$ and $\mathcal{F}^{-1}$ correspond to the discrete
Fourier transform and its inverse, respectively, and $\cdot$ denotes
element by element multiplication. Thus, it is in principle possible
to solve the KDI numerically for arbitrary lens shapes using the Fast
Fourier Transform (FFT). The technique is applied for plasma lenses
in one dimension by \citet{Watson&Melrose2006} and \citet{Melrose&Watson2006},
and in two dimensions by \citet{Stinebringetal2007} using code developed
by \citet{Colesetal1995}, and we use it in the main text to show
that it is possible to use an enhanced version of geometric optics
to reproduce the intensity fluctuations predicted by wave optics.

\begin{figure}[t]
\begin{centering}
\includegraphics[width=0.95\textwidth]{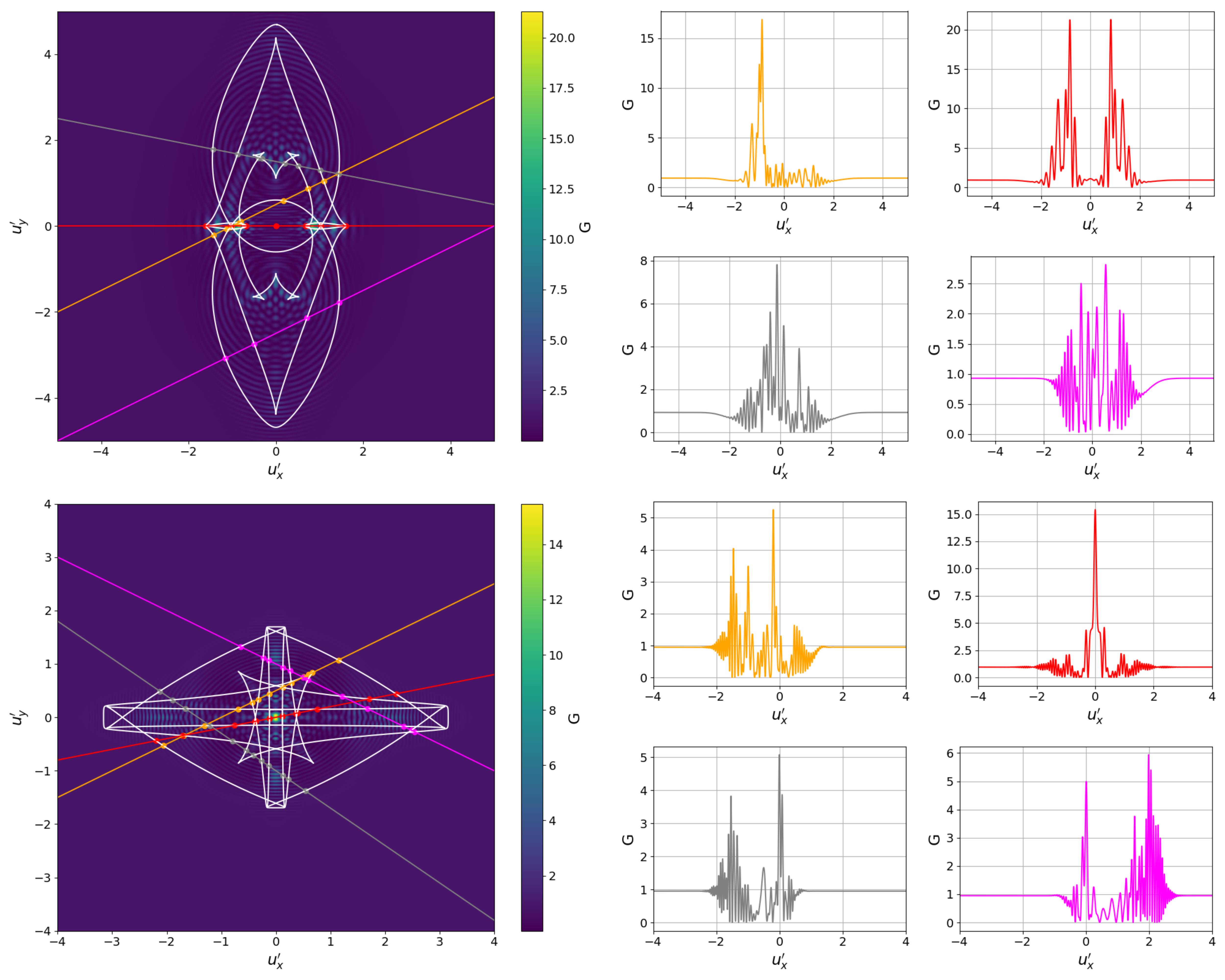}
\par\end{centering}
\caption{\emph{\label{fig: A1}Left}: Colormap of the gain in the $u'$ plane
overlaid with the caustic curves in white and slices along the plane
in different colors for two different lens shapes. The points of intersection
between the slices and the caustic lines are marked by points. The
top panel corresponds to a lens with shape $\psi(\boldsymbol{u})=0.74\left(u_{x}^{2}+u_{y}^{6}\right)\exp\left(-u_{x}^{2}-u_{y}^{2}\right)$,
and parameters $a_{x}=a_{y}=0.02$ AU, ${\rm DM}_{\ell}=-1.5\times10^{-6}$
pc cm\protect\textsuperscript{-3}, $\nu=0.8$ GHz, $d_{so}=1$ kpc,
and $d_{sl}=0.5$ kpc. The bottom lens has $\psi(\boldsymbol{u})=\exp\left(-u_{x}^{4}-u_{y}^{4}\right)$,
$a_{x}=0.04$ AU, $a_{y}=0.05$ AU, ${\rm DM}_{\ell}=-2\times10^{-6}$
pc cm\protect\textsuperscript{-3}, $\nu=1.0$ GHz, $d_{so}=5$ kpc,
$d_{sl}=2.5$ kpc. In both cases,\emph{ $\phi_{0}\approx50$ }rad\emph{.
Right}: Plots of the gain along the paths shown in the left panel
for each lens. Both kinds of lens show folds, cusps, and higher order
catastrophes. The top lens can generate up to nine images of the source,
whereas the bottom lens can produce up to seventeen.}
\end{figure}

Although useful, this approach suffers from serious limitations. First,
it does not give information about the number of images of the source
that can potentially be seen by the observer or the respective amplifications,
phases, and TOAs of each of these images. Second, in practice the
method can only be applied for a restricted range of lens scales $a_{x,y}$
and relatively small values of $\phi_{o}$. The issue is the grid
size necessary to properly sample the oscillations of the functions
$G(\boldsymbol{u},\nu)$ and $H(\boldsymbol{u},\nu)$. We illustrate
this for the former case. Consider a lens with characteristic scales
$a_{x}=a_{y}=a$. By Nyquist's sampling theorem, the maximum array
index $n_{max}$ that can be sampled along a given axis is given by
(\citealt{Schmidt2010})
\begin{equation}
n_{max}=\frac{\pi r_{F}^{2}}{(\Delta x)^{2}},
\end{equation}
where $\Delta x$ is the grid spacing in physical units. Now, let
$u_{max}'$ be the half-width of the $u'$ plane along either of the
axes, and $N$ be the size of the array along that axis. Then, the
sampling interval can be written as
\begin{equation}
\Delta x=\frac{2au_{max}'}{N}.
\end{equation}

Setting $N=n_{max}$ and rearranging, we have that the size of the
grid along one axis required to ensure proper sampling is
\begin{equation}
N=\frac{4a^{2}(u_{max}')^{2}}{\pi r_{F}^{2}}.
\end{equation}

This means that if we want to properly calculate the field for a lens
with size $a=1$ AU up to $u_{max}'=5$ and with distances $d_{sl}=0.5$
kpc, $d_{so}=1$ kpc, and frequency of observation $\nu=0.8$ GHz,
we need $N\approx1.5\times10^{6}$. This might be acceptable for the
one dimensional case, but a two dimensional grid with side of size
$N$ is too big for even a modern desktop computer to handle. A more
detailed analysis of sampling constraints and the numerical simulation
of wave propagation using Fourier optics can be found in \citet{Schmidt2010}.

Perhaps the primary advantage of this numerical strategy is that it
does not have any problem calculating the field at caustic regions
for any kind of catastrophe, even the higher order ones. Figure \ref{fig: A1}
shows the gain obtained using this method for different paths through
the $u'$ plane for a lens that shows higher order catastrophes than
the ones in the main text.

\section{Estimation of the value of the gain at the caustic}\label{sec: Appendix B}

For very large values of $\phi_{0}$, it might be desirable in some
cases to find the gain due to a lens using zeroth order geometric
optics (Eq. \ref{eq: 16}), since the oscillations due to multiple
imaging will give a value of the flux consistent with the prediction
from that equation once we take into account the frequency resolution
of the observations. Close to the caustics, however, the gain diverges.
When $\phi_{o}$ is large and the lens has strong refractive power,
the gain can diverge in such a way that the maximum value occurs extremely
close to the caustic, and this value can be estimated by the an extension
of the method of stationary phase.

This estimate has been derived in more than one dimension by just
a handful of authors in the context of asymptotic expansions of integrals,
and their results do not necessarily agree with each other. Here we
give two of the published formulas, specifically applied to the KDI,
although we do not derive them. According to \citet{Chako1965} and
\citet{Wong2001}, the gain at the singularity is
\begin{equation}
G_{max}=\frac{a_{x}^{2}a_{y}^{2}}{12\pi r_{F}^{4}}\frac{\Gamma^{2}(\nicefrac{1}{3})}{\left|\Phi_{20}\right|\left|\Phi_{03}\right|^{\nicefrac{2}{3}}}.
\end{equation}

\citet{Bleistein&Handelsman1975} and \citet{Cooke1982} give a more
complicated expression,
\begin{equation}
G_{max}=\frac{a_{x}^{2}a_{y}^{2}}{4\pi^{2}r_{F}^{4}}\left|\Phi_{20}\right|\Gamma^{2}(\nicefrac{1}{3})\left(\frac{32\pi^{2}}{3\left|B\right|^{2}}\right)^{\nicefrac{1}{3}},
\end{equation}
where $B=\Phi_{20}^{3}\Phi_{03}-3\Phi_{20}^{2}\Phi_{11}\Phi_{12}+3\Phi_{20}\Phi_{11}^{2}\Phi_{21}-\Phi_{11}^{3}\Phi_{30}$.
All derivatives of the phase in both equations are evaluated at the
degenerate stationary phase point for which $\Phi_{10}=\Phi_{01}=\Phi_{20}\Phi_{02}-\Phi_{11}^{2}=0$
. Some numerical experimentation has determined that both formulas
give similar but not the same results.

\section{Numerics}\label{sec: Appendix C}

The key behind successful application of geometric optics as presented
in the main text is the ability to numerically solve the lens equation,
Eq. \ref{eq: 14}. This is essentially a two dimensional nonlinear
root finding problem, with the added difficulties that the number
of roots can be more than one, and that roots can appear or disappear
as a function of the input parameters. A general method for two dimensional
root finding consists in rewriting the system of equations in the
form
\begin{equation}
\left[\begin{array}{c}
f(x,y)\\
g(x,y)
\end{array}\right]=\left[\begin{array}{c}
0\\
0
\end{array}\right],
\end{equation}
where $f(x,y)$ and $g(x,y)$ are the two equations that must be solved
simultaneously. Once this is done, we can produce contour plots of
both equations in order to find the sets of curves that satisfy $f(x,y)=0$
and $g(x,y)=0$. The roots of the two dimensional system will then
correspond to the points of intersection of these sets of curves.
When implemented properly, this method allows one to find all the
roots of a two dimensional system within a range of values for $x$
and $y$. A similar idea was pursued by \citet{Schramm&Kayser1987}
to solve the lens equation for gravitational lensing. The disadvantage
of this scheme is that it requires the evaluation of both $f(x,y)$
and $g(x,y)$ in a two dimensional grid that spans the area in which
we are looking for solutions, which can be very computationally expensive
if done repeatedly. 

Since we are interested in solving the equation at many different
points in parameter space, it is desirable to find a way to solve
the lens equation that does not require us to apply the above algorithm
at every single point of the independent variable. We can do this
by combining it with other, more efficient numerical techniques that
have been developed for numerical root finding in an arbitrary number
of dimensions. These have existed for a long time, and are available
for a variety of programming languages. In Python, some of these routines
are available via the SciPy\footnote{Jones E, Oliphant E, Peterson P, et al. SciPy: Open Source Scientific
Tools for Python, 2001-, http://www.scipy.org/ } library's optimization package. Although more efficient, these algorithms
have the limitation that they rely on the user to input a guess solution
that must be close enough to the actual solution. Furthermore, if
there are multiple roots, they will only find the one closest to the
input guess. This means that there is no way to find out exactly how
many roots there are for a particular set of parameters.

Our strategy consists in combining the contour plotting method with
the optimization algorithms in SciPy. First, we find the caustic locations
for the range of parameters that we want to find the solutions of
the lens equation for. If we are looking for solutions as a function
of $\boldsymbol{u'}$, we apply the contour plotting algorithm to
find the intersections between the curves in the $u'$ plane that
satisfy Eq. \ref{eq: 25} with the line $u_{y}'=mu_{x}'+n$, where
$m$ and $n$ parameterize the path through the $u'$ plane. If we
are looking for the solutions as a function of $\nu$, we apply the
contour plotting algorithm to simultaneously solve the system of equations
given in Eq. \ref{eq: 26}.

This step allows us to separate the regions in parameter space that
contain different numbers of solutions to the lens equation. Now,
we can apply the contour plotting method again to find the number
of roots at the center of each region. This results in the method
being more reliable, because close to region boundaries, at least
two roots will be very close to each other, whereas they will be maximally
separated at the region's center. After having found the roots at
the center of each of these regions, we find the other roots by iterating
forward and backward in parameter space, using the root finding algorithm
from SciPy with the previously found roots as the input guess solutions.
As long as the distance between neighboring values of the independent
variable is small enough, this strategy tends to work. It has the
advantage of being much more efficient than applying the contour plotting
method repeatedly, and also allows us to find the roots up to a very
close distance to the caustic. This method has been tried for a wide
variety of lens shapes and parameters, and has been found to be very
reliable, especially for finding the real roots of the lens equation.

In order to find the complex rays, we need to apply a modified version
of the above procedure that does not rely on contour plotting. The
reason is that extending the search of solutions to the complex plane
transforms the two dimensional lens equation into a four dimensional
equation, and evaluating four different equations in four dimensional
space is not practically feasible. Instead, we exploit the fact that,
as discussed in the main text, very close to the shadow side of a
caustic, the only important set of complex conjugate solutions to
the lens equation is the one that has the smallest magnitude of its
imaginary part. The real part of this complex conjugate set will be
almost the same as that of the solution to the lens equation that
intersects with the singularity. Thus, we use SciPy's root finding
algorithm with a value of the independent variable that falls in the
caustic's shadow side but at the same time is very close to the singularity,
and input the value of $u$ at the caustic as the guess solution.
From there, we can recursively look for complex solutions that are
farther away from the caustic in the same manner as we did for the
case of the real solutions.

\section{More numerical examples and lens colormaps}\label{sec: Appendix D}

\begin{figure}[H]
\begin{centering}
\includegraphics[width=0.9\textwidth]{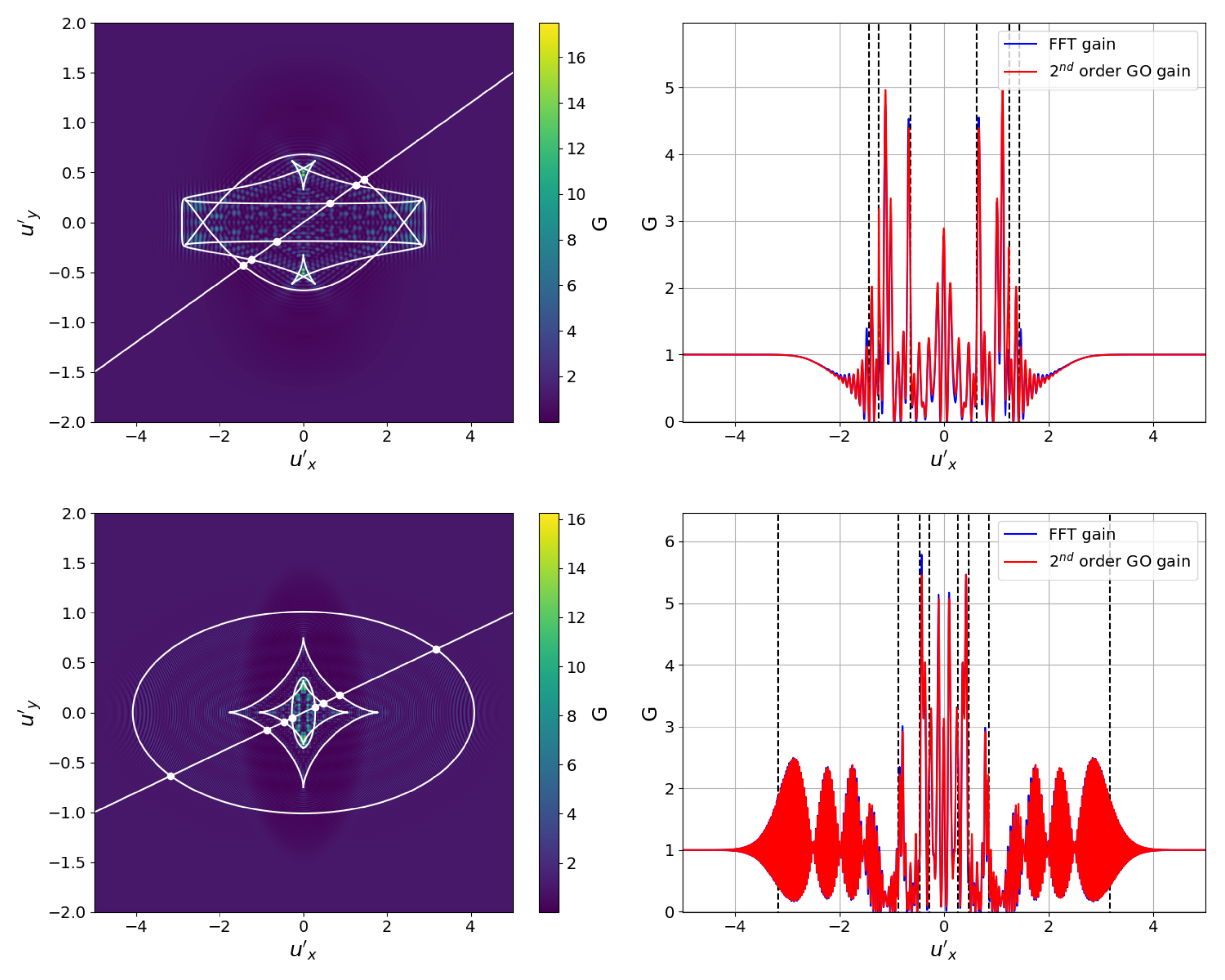}
\par\end{centering}
\caption{Comparison of the gains obtained from a full numerical solution of
the KDI and second order geometric optics. The left column shows color
maps of the gain obtained by solving the KDI via the FFT. The white
circles correspond to caustic curves, and the straight white line
shows the path of the observer through the $u'$ plane. The right
column shows the gain along this path as calculated via the FFT method
and second order geometric optics. The points of intersection between
the caustics and the observer path are marked by white points in the
left column and by dashed vertical black lines on the plots in the
right column. The top panel shows an underdense rectangular Gaussian
lens with $\psi(\boldsymbol{u})=\exp\left(-u_{x}^{2}-u_{y}^{4}\right)$,
lens phase $\phi_{0}=80$ rad, and lens scales $a_{x}=1.5\times10^{-2}$
AU and $a_{y}=3\times10^{-2}$ AU. The bottom panel corresponds to
an underense super-Gaussian lens with $\psi(\boldsymbol{u})=\exp\left[-\left(u_{x}^{2}+u_{y}^{2}\right)^{3}\right]$,
lens phase $\phi_{0}=120$ rad, and lens scales $a_{x}=2.5\times10^{-2}$
AU and $a_{y}=4\times10^{-2}$ AU. The frequency of observation is
$\nu=1.4$ GHz, $d_{so}=1$ kpc, $d_{sl}=0.5$ kpc for both the top
and bottom panels.}
 
\end{figure}

\newpage{}

\begin{figure}[H]
\begin{centering}
\includegraphics[width=0.95\textwidth]{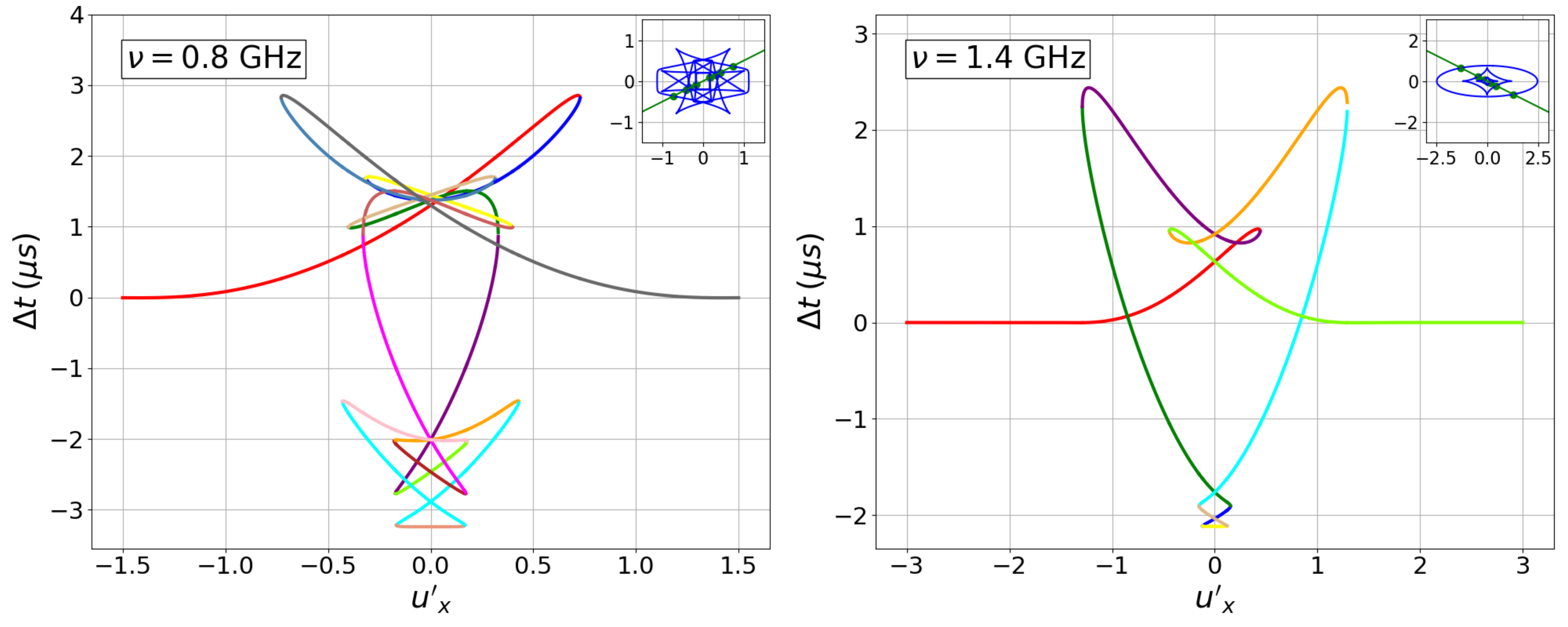}
\par\end{centering}
\caption{Individual image TOAs for two different lenses and different paths
through the u' plane. The left panel corresponds to a square Gaussian
lens with $\psi(\boldsymbol{u})=\exp\left(-u_{x}^{4}-u_{y}^{4}\right)$,
${\rm DM}_{\ell}=-5\times10^{-4}$ pc cm\protect\textsuperscript{-3},
$d_{so}=1$ kpc, $d_{sl}=0.5$ kpc, $a_{x}=0.5$ AU and $a_{y}=0.6$
AU. The frequency of observation is $\nu=0.8$ GHz, which gives a
lens phase of $\phi_{o}\approx1.63\times10^{4}$ rad. The right panel
corresponds to a super-Gaussian lens with $\psi(\boldsymbol{u})=\exp\left[-\left(u_{x}^{2}+u_{y}^{2}\right)^{2}\right]$,
${\rm DM}_{\ell}=-1\times10^{-3}$ pc cm\protect\textsuperscript{-3},
$d_{so}=5$ kpc, $d_{sl}=2.5$ kpc, $a_{x}=0.7$ AU and $a_{y}=1$
AU, with $\nu=1.4$ GHz, and thus $\phi_{o}\approx1.86\times10^{4}$
rad. Different colors denote different images, and the top right subplots
show the path of the observer through the $u'$ plane and the caustic
curves. The maximum number of images in each plot is seventeen and
nine, respectively.}
 
\end{figure}

\begin{figure}[H]
\begin{centering}
\includegraphics[width=0.95\textwidth]{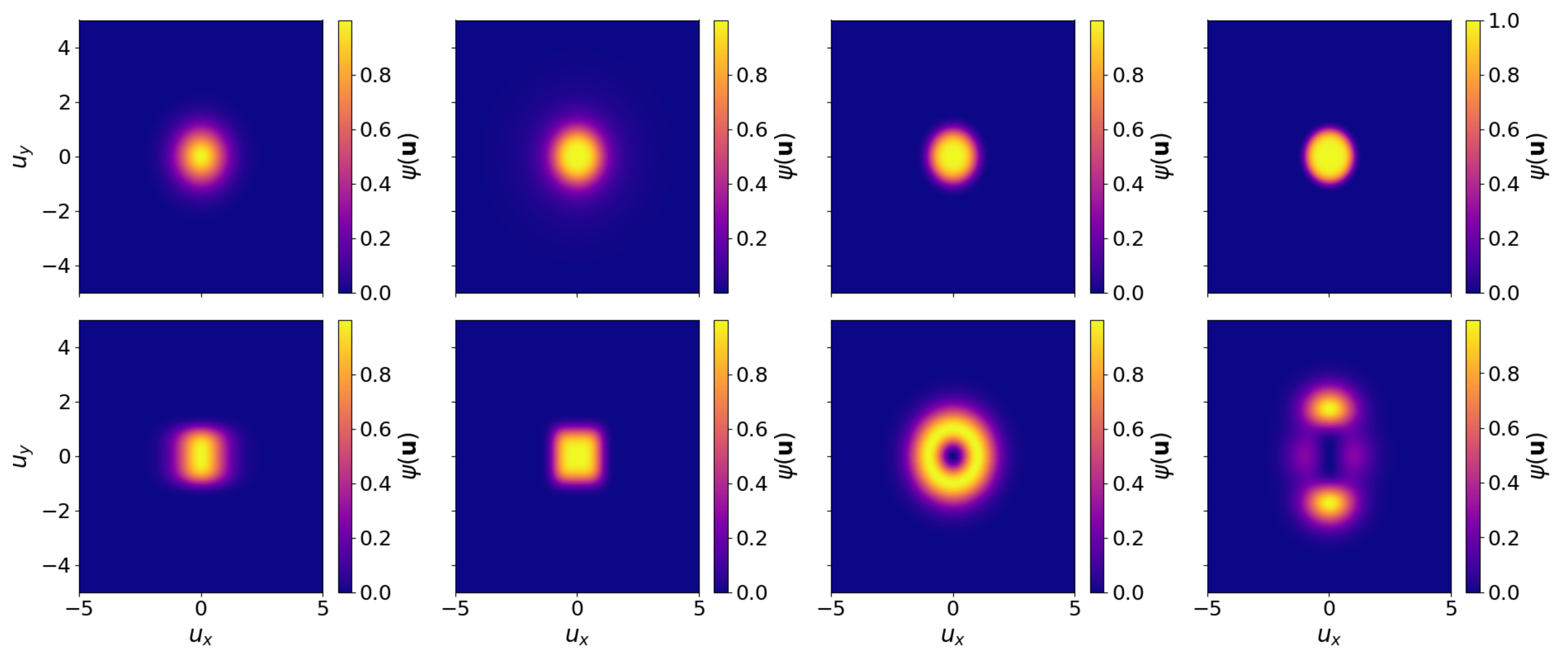}
\par\end{centering}
\caption{Colormaps of the different types of lensing structures used in the
text and appendices. \emph{Top row, from left to right}: Gaussian
lens, $\psi(\boldsymbol{u})=\exp\left(-u_{x}^{2}-u_{y}^{2}\right)$,
Lorentzian lens, $\psi(\boldsymbol{u})=\nicefrac{1}{\left[\left(u_{x}^{2}+u_{y}^{2}\right)^{2}+1\right]}$,
and super-Gaussian lenses $\psi(\boldsymbol{u})=\exp\left[-\left(u_{x}^{2}+u_{y}^{2}\right)^{2}\right]$
and $\psi(\boldsymbol{u})=\exp\left[-\left(u_{x}^{2}+u_{y}^{2}\right)^{3}\right]$.
\emph{Bottom row, from left to right}: Rectangular Gaussian lens,
$\psi(\boldsymbol{u})=\exp\left(-u_{x}^{2}-u_{y}^{4}\right)$, square
Gaussian lens, $\psi(\boldsymbol{u})=\exp\left(-u_{x}^{4}-u_{y}^{4}\right)$,
ring-like lens $\psi(\boldsymbol{u})=2.72\left(u_{x}^{2}+u_{y}^{2}\right)\exp\left(-u_{x}^{2}-u_{y}^{2}\right)$,
and double lens $\psi(\boldsymbol{u})=0.74\left(u_{x}^{2}+u_{y}^{6}\right)\exp\left(-u_{x}^{2}-u_{y}^{2}\right)$.}
 
\end{figure}

\newpage{}

\bibliographystyle{aasjournal}
\bibliography{fullbib}

\begin{thebibliography}{}
\expandafter\ifx\csname natexlab\endcsname\relax\def\natexlab#1{#1}\fi
\providecommand{\url}[1]{\href{#1}{#1}}
\providecommand{\dodoi}[1]{doi:~\href{http://doi.org/#1}{\nolinkurl{#1}}}
\providecommand{\doeprint}[1]{\href{http://ascl.net/#1}{\nolinkurl{http://ascl.net/#1}}}
\providecommand{\doarXiv}[1]{\href{https://arxiv.org/abs/#1}{\nolinkurl{https://arxiv.org/abs/#1}}}

\bibitem[{Bannister {et~al.}(2016)Bannister, Stevens, Tuntsov, Walker,
  Johnston, Reynolds, \& Bignall}]{Bannisteretal2016}
Bannister, K.~W., Stevens, J., Tuntsov, A.~V., {et~al.} 2016, Science, 351, 354

\bibitem[{Berry(1976)}]{Berry1976}
Berry, M.~V. 1976, AdPhy, 25, 1

\bibitem[{Berry \& Upstill(1980)}]{Berry&Upstill1980}
Berry, M.~V., \& Upstill, C. 1980, in Progress in Optics, ed. E.~Wolf, Vol.~18
  (Elsevier), 257--346

\bibitem[{Bleistein \& Handelsman(1975)}]{Bleistein&Handelsman1975}
Bleistein, N., \& Handelsman, R.~A. 1975, Asymptotic expansions of integrals
  (Dover Publications)

\bibitem[{{Born} \& {Wolf}(1999)}]{Born&Wolf1999}
{Born}, M., \& {Wolf}, E. 1999, {Principles of Optics}, seventh edn. (Cambridge
  University Press)

\bibitem[{Borovikov \& Kinber(1994)}]{Borovikov&Kinber1994}
Borovikov, V.~A., \& Kinber, B.~E. 1994, Geometrical theory of diffraction,
  Electromagnetic Wave Series No.~37 (The Institution of Electrical Engineers)

\bibitem[{{Budden} \& {Terry}(1971)}]{Budden&Terry1971}
{Budden}, K.~G., \& {Terry}, P.~D. 1971, RSPSA, 321, 275

\bibitem[{Chako(1965)}]{Chako1965}
Chako, N. 1965, JApMa, 1, 372

\bibitem[{Chester {et~al.}(1957)Chester, Friedman, \& Ursell}]{Chester1957}
Chester, C., Friedman, B., \& Ursell, F. 1957, 53, 599

\bibitem[{Clegg {et~al.}(1996)Clegg, Fey, \& Fiedler}]{Cleggetal1996}
Clegg, A.~W., Fey, A.~L., \& Fiedler, R.~L. 1996, ApJ, 457, L23

\bibitem[{{Clegg} {et~al.}(1998){Clegg}, {Fey}, \& {Lazio}}]{Cleggetal1998}
{Clegg}, A.~W., {Fey}, A.~L., \& {Lazio}, T.~J.~W. 1998, ApJ, 496, 253

\bibitem[{Cognard {et~al.}(1993)Cognard, Bourgois, Lestrade,
  {et~al.}}]{Cognardetal1993}
Cognard, I., Bourgois, G., Lestrade, J.-F., {et~al.} 1993, Nature, 366, 320

\bibitem[{Coles {et~al.}(1995)Coles, Filice, Frehlich, \&
  Yadlowsky}]{Colesetal1995}
Coles, W.~A., Filice, J.~P., Frehlich, R.~G., \& Yadlowsky, M. 1995, ApOpt, 34,
  2089

\bibitem[{Coles {et~al.}(2015)Coles, Kerr, Shannon, {et~al.}}]{Colesetal2015}
Coles, W.~A., Kerr, M., Shannon, R.~M., {et~al.} 2015, ApJ, 808, 113

\bibitem[{Connor(1973{\natexlab{a}})}]{Connor1973a}
Connor, J. 1973{\natexlab{a}}, MolPh, 25, 181

\bibitem[{Connor(1973{\natexlab{b}})}]{Connor1973b}
Connor, J. N.~L. 1973{\natexlab{b}}, MolPh, 26, 1217

\bibitem[{Cooke(1982)}]{Cooke1982}
Cooke, J.~C. 1982, JApMa, 29, 25

\bibitem[{Cordes {et~al.}(1986)Cordes, Pidwerbetsky, \&
  Lovelace}]{Cordesetal1986}
Cordes, J.~M., Pidwerbetsky, A., \& Lovelace, R.~E. 1986, ApJ, 310, 737

\bibitem[{Cordes {et~al.}(2006)Cordes, Rickett, Stinebring, \&
  Coles}]{Cordesetal2006}
Cordes, J.~M., Rickett, B.~J., Stinebring, D.~R., \& Coles, W.~A. 2006, ApJ,
  637, 346

\bibitem[{Cordes \& Shannon(2010)}]{Cordes&Shannon2010}
Cordes, J.~M., \& Shannon, R.~M. 2010.
\newblock \doarXiv{1010.3785}

\bibitem[{Cordes {et~al.}(2016)Cordes, Shannon, \& Stinebring}]{Cordesetal2016}
Cordes, J.~M., Shannon, R.~M., \& Stinebring, D.~R. 2016, ApJ, 817, 16

\bibitem[{Cordes \& Wasserman(2016)}]{Cordes&Wasserman2016}
Cordes, J.~M., \& Wasserman, I. 2016, MNRAS, 457, 232

\bibitem[{{Cordes} {et~al.}(2017){Cordes}, {Wasserman}, {Hessels}, {Lazio},
  {Chatterjee}, \& {Wharton}}]{Cordesetal2017}
{Cordes}, J.~M., {Wasserman}, I., {Hessels}, J.~W.~T., {et~al.} 2017, ApJ, 842,
  10

\bibitem[{Cordes \& Wolszcan(1986)}]{Cordes&Wolszcan1986}
Cordes, J.~M., \& Wolszcan, A. 1986, ApJ, 307, L27

\bibitem[{Dai \& Lu(2017)}]{Dai&Lu2017}
Dai, L., \& Lu, W. 2017, ApJ, 847, 19

\bibitem[{Er \& Rogers(2017)}]{Er&Rogers2017}
Er, X., \& Rogers, A. 2017, MNRAS, 475, 867

\bibitem[{Fiedler {et~al.}(1994)Fiedler, Dennison, Johnston,
  {et~al.}}]{Fiedleretal1994}
Fiedler, R., Dennison, B., Johnston, K.~J., {et~al.} 1994, ApJ, 430, 581

\bibitem[{{Fiedler} {et~al.}(1987){Fiedler}, {Dennison}, {Johnston}, \&
  {Hewish}}]{Fiedleretal1987}
{Fiedler}, R.~L., {Dennison}, B., {Johnston}, K.~J., \& {Hewish}, A. 1987,
  Nature, 326, 675

\bibitem[{{Goodman} {et~al.}(1987){Goodman}, {Romani}, {Blandford}, \&
  {Narayan}}]{Goodmanetal1987}
{Goodman}, J.~J., {Romani}, R.~W., {Blandford}, R.~D., \& {Narayan}, R. 1987,
  MNRAS, 229, 73

\bibitem[{Gupta {et~al.}(1999)Gupta, Ghat, \& Rao}]{Guptaetal1999}
Gupta, Y., Ghat, N. D.~R., \& Rao, A.~P. 1999, ApJ, 520, 173

\bibitem[{Gupta {et~al.}(1994)Gupta, Rickett, \& Lyne}]{Guptaetal1994}
Gupta, Y., Rickett, B.~J., \& Lyne, A.~G. 1994, MNRAS, 269, 1035

\bibitem[{Howls(1997)}]{Howls1997}
Howls, C.~J. 1997, RSPSA, 453, 2271

\bibitem[{Kaminski(1994)}]{Kaminsky1994}
Kaminski, D. 1994, MApAn, 1, 44

\bibitem[{Katsaounis {et~al.}(2001)Katsaounis, Kossioris, \&
  Makrakis}]{Katsaounisetal2001}
Katsaounis, T., Kossioris, G.~T., \& Makrakis, G.~N. 2001, Math. Models Meth.
  Appl. Sci., 11, 199

\bibitem[{Keith {et~al.}(2013)Keith, Coles, Shannon, Hobbs, Manchester, Bailes,
  Bhat, Burke-Spolaor, Champion, Chaudhary, {et~al.}}]{Keithetal2013}
Keith, M.~J., Coles, W.~A., Shannon, R.~M., {et~al.} 2013, MNRAS, 429, 2161

\bibitem[{Kerr {et~al.}(2017)Kerr, Coles, Ward, {et~al.}}]{Kerretal2017}
Kerr, M., Coles, W.~A., Ward, C.~A., {et~al.} 2017, MNRAS, 474, 4637

\bibitem[{Kravstov(1968)}]{Kravstov1968}
Kravstov, Y.~A. 1968, SPhAc, 14, 1

\bibitem[{Kravtsov {et~al.}(1999)Kravtsov, Forbes, \&
  Asatryan}]{Kravstovetal1999}
Kravtsov, Y.~A., Forbes, G.~W., \& Asatryan, A.~A. 1999, in Progress in optics,
  ed. E.~Wolf, Vol.~39 (Elsevier), 1--62

\bibitem[{Kravtsov \& Orlov(1999)}]{Kravstov&Orlov1999}
Kravtsov, Y.~A., \& Orlov, Y.~I. 1999, Springer Series on Wave Phenomena,
  Vol.~15, Caustics, catastrophes and wave fields, 2nd edn. (Springer)

\bibitem[{Kryukovskii {et~al.}(2006)Kryukovskii, Lukin, Palkin, \&
  Rastyagaev}]{Kryukovskiietal2006}
Kryukovskii, A.~S., Lukin, D.~S., Palkin, E.~A., \& Rastyagaev, D.~S. 2006,
  Journal of Communications Technology and Electronics, 51, 1087

\bibitem[{Lam {et~al.}(2018)Lam, Ellis, Grillo, {et~al.}}]{Lametal2018}
Lam, M.~T., Ellis, J.~A., Grillo, G., {et~al.} 2018, ApJ, 861, 132

\bibitem[{Ludwig(1966)}]{Ludwig1966}
Ludwig, D. 1966, Commun. Pure Appl. Math., 19, 215

\bibitem[{Main {et~al.}(2018)Main, Yang, Chan, Li, Lin, Mahajan, Pen,
  Vanderlinde, \& Kerkwijk}]{Mainetal2018}
Main, R., Yang, I.~S., Chan, V., {et~al.} 2018, Nature, 557, 522

\bibitem[{{Melrose} \& {Watson}(2006)}]{Melrose&Watson2006}
{Melrose}, D.~B., \& {Watson}, P.~G. 2006, ApJ, 647, 1131

\bibitem[{Nye(1978)}]{Nye1978}
Nye, J.~F. 1978, RSPSA, 361, 21

\bibitem[{Paris(1991)}]{Paris1991}
Paris, R.~B. 1991, RSPSA, 432, 391

\bibitem[{Pearcey(1946)}]{Pearcey1946}
Pearcey, T. 1946, PMag, 37, 311

\bibitem[{Pen \& King(2012)}]{Pen&King2012}
Pen, U.~L., \& King, L. 2012, MNRAS, 421, L132

\bibitem[{Poston \& Stewart(1978)}]{Poston&Stewart1978}
Poston, T., \& Stewart, I. 1978, Catastrophe Theory and its Applications (Dover
  Publications)

\bibitem[{Pushkarev {et~al.}(2013)Pushkarev, Kovalev, Lister,
  {et~al.}}]{Pushkarevetal2013}
Pushkarev, A.~B., Kovalev, Y.~Y., Lister, M.~L., {et~al.} 2013, A\&A, 555, A80

\bibitem[{{Qiu} \& {Wong}(2000)}]{Qiu&Wong2000}
{Qiu}, W.-Y., \& {Wong}, R. 2000, RSPSA, 456, 407

\bibitem[{Rickett(1990)}]{Rickett1990}
Rickett, B.~J. 1990, ARA\&A, 28, 561

\bibitem[{Schmidt(2010)}]{Schmidt2010}
Schmidt, J.~D. 2010, Numerical simulation of optical wave propagation with
  examples in MATLAB

\bibitem[{Schneider {et~al.}(1992)Schneider, Ehlers, \&
  Falco}]{Schneideretal1992}
Schneider, P., Ehlers, J., \& Falco, E.~E. 1992, {Gravitational Lenses}
  (Springer)

\bibitem[{{Schramm} \& {Kayser}(1987)}]{Schramm&Kayser1987}
{Schramm}, T., \& {Kayser}, R. 1987, A\&A, 174, 361

\bibitem[{Schramm \& Kayser(1995)}]{Schramm&Kayser1995}
Schramm, T., \& Kayser, R. 1995, A\&A, 299, 1

\bibitem[{Stamnes(1986)}]{Stamnes1986}
Stamnes, J.~J. 1986, Waves in Focal Regions: Propagation, Diffraction and
  Focusing of Light, Sound and Water Waves (Adam Hilger)

\bibitem[{{Stinebring} {et~al.}(2007){Stinebring}, {Matters}, \&
  {Hemberger}}]{Stinebringetal2007}
{Stinebring}, D., {Matters}, J., \& {Hemberger}, D. 2007, in Astronomical
  Society of the Pacific Conference Series, Vol. 365, SINS - Small Ionized and
  Neutral Structures in the Diffuse Interstellar Medium, ed. M.~{Haverkorn} \&
  W.~M. {Goss}, 275

\bibitem[{Thom(1972)}]{Thom1972}
Thom, R. 1972, Stabilite structurelle et morphogenese (Benjamin)

\bibitem[{Thorne \& Blandford(2017)}]{Thorne&Blandford2017}
Thorne, K.~S., \& Blandford, R.~D. 2017, Modern Classical Physics (Princeton
  University Press)

\bibitem[{Tuntsov {et~al.}(2016)Tuntsov, Walker, Koopmans, Bannister, Stevens,
  Johnston, Reynolds, \& Bignall}]{Tuntsovetal2016}
Tuntsov, A.~V., Walker, M.~A., Koopmans, L. V.~E., {et~al.} 2016, ApJ, 817, 176

\bibitem[{Ursell(1965)}]{Ursell1965}
Ursell, F. 1965, PCPS, 61, 113

\bibitem[{Vedantham {et~al.}(2017)Vedantham, Readhead, Hovatta,
  {et~al.}}]{Vedanthametal2017}
Vedantham, H.~K., Readhead, A. C.~S., Hovatta, T., {et~al.} 2017, ApJ, 845, 90

\bibitem[{{Watson} \& {Melrose}(2006)}]{Watson&Melrose2006}
{Watson}, P.~G., \& {Melrose}, D.~B. 2006, ApJ, 647, 1142

\bibitem[{Wong(2001)}]{Wong2001}
Wong, R. 2001, Asymptotic approximations of integrals, Vol.~34 (SIAM)

\end{thebibliography}

\end{document}